\documentclass[fleqn,usenatbib]{mnras}

\usepackage[T1]{fontenc}
\usepackage{ae,aecompl}
\usepackage{bm}

\usepackage{graphicx}	
\usepackage{amsmath}	
\usepackage{amssymb}	
\usepackage{mathrsfs}
\usepackage{lipsum}
\usepackage{hyperref}
\usepackage[dvipsnames]{xcolor}
\usepackage{newtxtext,newtxmath}
\usepackage{rotating}
\usepackage{blindtext}
\usepackage{multicol}
\usepackage{multirow}

\newcommand{\LCDM}{$\Lambda$CDM }
\newcommand{\Msun}{\, \rm M_{\odot}}

\title[Impact of baryons on halo structure]{The impact of baryons on the internal structure of dark matter haloes from dwarf galaxies to superclusters in the redshift range $0<z<7$}

\author[D. Sorini et al.]{%
Daniele Sorini$^{1,}$\thanks{E-mail: daniele.sorini@durham.ac.uk}, 
Sownak Bose$^1$, 
R\"udiger Pakmor$^{2}$, 
Lars Hernquist$^{3}$, 
Volker Springel$^{2}$,
\newauthor
Boryana Hadzhiyska$^{4,\, 5}$, 
C\'esar Hern\'andez-Aguayo$^{2,\, 6}$ and 
Rahul Kannan$^{7}$
\\%
$^1$Institute for Computational Cosmology, Department of Physics, Durham University, South Road, Durham, DH1 3LE, United Kingdom\\%
$^{2}$Max-Planck-Institut f\"ur Astrophysik, Karl-Schwarzschild-Str. 1, D-85748, Garching, Germany\\%
$^{3}$Center for Astrophysics | Harvard \& Smithsonian, 60 Garden St, Cambridge, MA 02138, USA\\%
$^{4}$Physics Division, Lawrence Berkeley National Laboratory, Berkeley, CA 94720, USA\\%
$^{5}$Berkeley Center for Cosmological Physics, Department of Physics, University of California, Berkeley, CA 94720, USA\\%
$^{6}$Excellence Cluster ORIGINS, Boltzmannstrasse 2, D-85748 Garching, Germany\\%
$^{7}$Department of Physics and Astronomy, York University, 4700 Keele Street, Toronto, ON M3J 1P3, Canada
}

\pubyear{2024}

\begin{document}
\label{firstpage}
\pagerange{\pageref{firstpage}--\pageref{lastpage}}
\maketitle

\begin{abstract}
We investigate the redshift evolution of the concentration-mass relationship of dark matter haloes in state-of-the-art cosmological hydrodynamic simulations and their dark-matter-only counterparts. By combining the IllustrisTNG suite and the novel MillenniumTNG simulation, our analysis encompasses a wide range of box size ($50 - 740 \: \rm cMpc$) and mass resolution ($8.5 \times 10^4 - 3.1 \times 10^7 \: \rm M_{\odot}$ per baryonic mass element). This enables us to study the impact of baryons on the concentration-mass relationship in the redshift interval $0<z<7$ over an unprecedented halo mass range, extending from dwarf galaxies to superclusters ($\sim 10^{9.5}-10^{15.5} \, \rm M_{\odot}$). We find that the presence of baryons increases the steepness of the concentration-mass relationship at higher redshift, and demonstrate that this is driven by adiabatic contraction of the profile, due to gas accretion at early times, which promotes star formation in the inner regions of haloes. At lower redshift, when the effects of feedback start to become important, baryons decrease the concentration of haloes below the mass scale $\sim 10^{11.5} \, \rm M_{\odot}$. Through a rigorous information criterion test, we show that broken power-law models accurately represent the redshift evolution of the concentration-mass relationship, and of the relative difference in the total mass of haloes induced by the presence of baryons. We provide the best-fit parameters of our empirical formulae, enabling their application to models that mimic baryonic effects in dark-matter-only simulations over six decades in halo mass in the redshift range $0<z<7$.
\end{abstract}




\section{Introduction}

Understanding how galaxy formation unfolds throughout the history of the Universe is a fundamental question that lies at the crossroads of galactic astrophysics and cosmology. The two key elements shaping the buildup of galaxies in a cosmological context are the hierarchical structure formation of dark matter (DM) haloes, and the astrophysical processes that shape star formation and the gaseous environment of galaxies. 

The former question is well understood within the standard \LCDM paradigm, thanks to early analytical models for the formation of DM haloes via hierarchical merging \citep{LaceyCole1993}, and N-body cosmological simulations following the evolution of self-gravitating DM particles \citep{Springel_2005, Klypin_2011, Angulo_2012, Fosalba_2015}. Given that DM haloes constitute the backbone within which galaxies form, understanding their internal structure represents a stepping stone towards a complete theory for cosmological galaxy formation.

A key result from early N-body simulations is that the spherically averaged DM density distribution, $\rho(r)$, within galactic haloes can be universally described by the so-called Navarro-Frenk-White (NFW) profile \citep{NFW}:
\begin{equation}
   \label{eq:NFW_rs}
      \frac{\rho(r)}{\rho_{\rm c}} = \frac{\Delta_{\rm c}}{\frac{r}{r_s}\left(1 + \frac{r}{r_s}\right)^2} \, ,
\end{equation}
where $\rho_{\rm c}$ is the critical density of the Universe. The `scale radius' $r_{\rm s}$ is a free parameter representing how concentrated the matter distribution is towards the centre of the halo. In fact, equation~\eqref{eq:NFW_rs} is often written in terms of the `concentration' parameter, defined as $c_{\rm 200c}=r_{\rm 200 c}/r_{\rm s}$, where $r_{\rm 200c}$ is the halocentric distance enclosing a total mass density equal to 200 times the critical density of the Universe, and is usually adopted as a proxy for the virial radius:
\begin{equation}
   \label{eq:NFW}
      \frac{\rho(r)}{\rho_{\rm c}} = \frac{\Delta_{\rm c}}{c_{\rm 200c} \frac{r}{r_{\rm 200c}}\left(1 + c_{\rm 200c}  \frac{r}{r_{\rm 200c}}\right)^2} \, .
\end{equation}
The parameter $\Delta_{\rm c}$ then regulates the normalisation of the profile such that its integral over the volume of the halo matches the virial mass. It follows that $\Delta_{\rm c}$ depends on the concentration, which is thus the only free parameter of the NFW profile.

Later studies suggested that the DM density distribution within haloes can be better characterised by incorporating an extra `shape' parameter \citep{Navarro_2010}, which appears in other frequently employed models, such as the Einasto profile \citep{Einasto, Merritt_2006}. But regardless of the specific functional form, DM density profiles still display a certain level of universality within N-body simulations, and the concentration remains a key parameter in the description of the halo structure. If the relationship between concentration and total mass of haloes is known, then the DM density profile of any halo of a given mass can be straightforwardly predicted. Thus, several cosmological N-body simulations tested the validity of the NFW or Einasto profile and investigated the concentration-mass relationship \citep{Bullock_2001, Maccio_2007, Neto_2007, Duffy_2008, Maccio_2008, Bhattacharya_2013, Dutton_2014, Ludlow_2014, Klypin_2016}, either within the \LCDM model or beyond \citep[e.g.][]{Bose_2016, Ruan_2023}. The halo mass and redshift range probed were progressively expanded with the advancement of numerical techniques and computational facilities.

For instance, using a set of nested zoom-in N-body simulations, \cite{Wang_2020} obtained the present-day concentration-mass relationship over 20 orders of magnitude in the halo mass range ($10^{-6} - 10^{14} \, \rm M_{\odot}$), hence verifying the robustness of the Einasto profile as a model for the DM distribution within collapsed structures. More recently, \cite{Uchuu} utilised the large-volume ($2.0 \, h^{-1} \, \rm Gpc$) and high-resolution ($8.97 \times 10^5\, h^{-1}\,\rm M_{\odot}$) Uchuu and Shin-Uchuu cosmological N-body simulations to probe the NFW concentration-mass relationship in the halo mass range $10^9 - 10^{15} \, \rm M_{\odot}$, studying its evolution in the redshift range $0<z<7$. All works consistently confirmed a decreasing concentration-mass relationship at lower redshift, proposing either a power-law fitting function \citep[e.g.][]{Dutton_2014, Schaller_2015}, or more complex, physically motivated analytical models \citep[e.g.][]{Ludlow_2013, Ludlow_2014, Ludlow_2016, Diemer_2019} following the evolution of collapsed structures. Other studies sought to directly connect the DM density profiles of haloes to large-scale structure statistics such as the power spectrum of density perturbations \citep{Diemer_2019, Brown_2020, Brown_2022}. 

The near universality of the DM density profiles in N-body simulations (at least in relaxed haloes; see e.g. discussion in \citealt{Diemer_2019}) descends from the scale-free behaviour induced by gravity. However, this does not hold true once baryons are included, as baryon-driven astrophysical processes introduce new characteristic scales that break the self-similarity of the DM density profiles. For example, gas cooling and dissipation \citep{White_1978, White_1991}, combined with the subsequent star formation, can alter the structure of the halo. The early adiabatic contraction model suggested that baryon collapse would increase the density of haloes in their central region \citep{Blumenthal_1986}. However, this model was found to overpredict the increase of DM density in hydrodynamic cosmological simulations \citep{Gnedin_2004, Gustaffson_2006}. Idealised simulations including a simplified outflow model reached qualitatively different conclusions, generating haloes with a central core \citep{Navarro_1996}. While several cosmological simulations confirmed this result \citep{Dehnen_2005, Read_2005, Mashchenko_2006, Governato_2010, Pontzen_2012, Martizzi_2013, Teyssier_2013}, others highlighted that the formation of cores in dwarf galaxies is either not ubiquitous \citep{Onorbe_2015} or outright absent \citep{Bose_2019}.

The development of more sophisticated cosmological hydrodynamic simulations, following the co-evolution of several species of baryonic matter, such as gas, stars, and black holes \citep[e.g.][]{Dolag_2009, OWLS, Dubois_2014, Illustris_V2014, Lukic_2015, EAGLE_Schaye2015, Simba_Dave2019, Flamingo}, expanded the scope of the inquiry. Indeed, different simulations rely on a variety of numerical prescriptions for sub-grid processes such as outflows driven by stars or active galactic nuclei (AGN; see e.g. \citealt{feedback_review} for a review). This prompts the question of how individual stellar and AGN feedback models, and not only the mere presence of baryons, affect the properties of galaxies and their host DM haloes. In this respect, understanding the impact of baryonic physics on the matter content and distribution within haloes remains a central question. 

\cite{Schaller_2015} showed that the spherically averaged DM density distribution within haloes is well represented by an NFW profile both in the EAGLE hydrodynamic cosmological simulation \citep{EAGLE_Schaye2015} and in its dark-matter-only (DMO) counterpart. The average concentration-mass relationship at $z=0$ was fit with a power law in both runs, and the hydrodynamic version exhibited a larger normalisation and gentler slope than the DMO variant. Using the same simulations, but applying different analysis techniques, \cite{Beltz-Mohrmann_2021} reached similar conclusions regarding the slope (but not the normalisation) of the relationship. The same work additionally considered the Illustris \citep{Illustris_V2014} and IllustrisTNG \citep{IllustrisTNG2018} hydrodynamic simulations. The former produced a steeper concentration-mass relationship with respect to its DMO variant, while the latter exhibited the opposite trend. Other studies focused on modelling the redshift evolution of the concentration-mass relationship in hydrodynamic simulations, rather than making comparisons with DMO runs \citep{Shirasaki_2018, Ragagnin_2019, Ragagnin_2021}, showing that the concentration of haloes of a fixed mass increases at later times. More recently, \cite{Shao_2023} used the CAMELS suite of simulations \citep{Camels} to show that the concentration-mass relationship in IllustrisTNG-type models of galaxy formation \citep{Weinberger_2017, IllustrisTNG2018} exhibits a plateau in the mass range $10^{11} - 10^{11.5} \, \rm M_{\odot}$. Instead, such a feature is absent in CAMELS boxes incorporating prescriptions based on the Simba \citep{Simba_Dave2019} cosmological simulations \citep[see also][]{Shao_2024}. Thus, all aforementioned works confirm that the concentration-mass relationship can change both qualitatively and quantitatively depending on the galaxy formation model embedded in cosmological simulations. 

A challenge in any analysis involving hydrodynamic simulations is the heavy computational cost, which imposes a trade-off between box size and mass resolution. Such numerical constraints translate into upper and lower limits on the halo mass range that can be probed. Combining three variants of the IllustrisTNG simulations with box size ranging from approximately $50 \, \rm Mpc$ to $300 \, \rm Mpc$, and mass resolution as good as $\sim 4.5 \times 10^5 \, \rm M_{\odot}$, \cite{Anbajagane_2022} managed to study the present-day concentration-mass relationship for halo masses between $\sim 10^9 \, \rm M_{\odot}$ and $10^{14.5}\, \rm M_{\odot}$. This constitutes an improvement of at least one order of magnitude with respect to the previously mentioned studies with hydrodynamic simulations. 

\begin{table*}
    \centering
    \caption{Properties of the simulations utilised for the primary analysis in this work. From left to right, the columns report: the type of simulation (hydrodynamic/DMO); the parent project (IllustrisTNG/MillenniumTNG); the simulation label; the box size; the number of DM particles; the number of initial gas elements; the mass of each DM particle; the average mass of the initial gas elements; the gravitational softening length for DM and (for the hydrodynamic simulations) stars; the minimum gravitational softening length for gas elements. The runs utilised for the main analysis are indicated in boldface. All other runs are reserved exclusively for convergence tests (see Appendix~\ref{app:convergence}).}
    \label{tab:simulations}
    \begin{tabular}{cccccccccc}
        \hline
        Type & Project & Name & Box size & $N_{\mathrm{DM}}$ & $N_{\mathrm{gas}}$ & $m_{\mathrm{DM}}$ & $m_{\mathrm{gas}}$ & $\varepsilon_{\mathrm{DM}, \star}$ & $\varepsilon_{\rm gas, \, min}$\\
         & &  & $\rm [cMpc]$ & &  & $[\rm M_{\odot}]$ & $[\rm M_{\odot}]$ & $[\rm kpc]$ & $[\rm pc]$ \\
        \hline
        Hydrodynamic & MillenniumTNG & \textbf{MTNG-740} & \textbf{740} & $\mathbf{4320^3}$ & $\mathbf{4320^3}$ & $\mathbf{1.65 \times 10^8}$ & $\mathbf{2.95 \times 10^7}$ & \textbf{3.7} & \textbf{370}\\
        & & MTNG-185 & 185 & $1080^3$ & $1080^3$ & $1.65 \times 10^8$ & $2.95 \times 10^7$ & 3.7 & 370\\
        & & MTNG-93 & 93 & $540^3$ & $540^3$ & $1.65 \times 10^8$ & $2.95 \times 10^7$ & 3.7 & 370\\
         & IllustrisTNG & \textbf{TNG-300} & \textbf{302.6} & $\mathbf{2500^3}$ & $\mathbf{2500^3}$ & $\mathbf{5.9 \times 10^7}$ & $\mathbf{1.1 \times 10^7}$ & \textbf{1.48} & \textbf{185} \\
         & & TNG-300-2 & 302.6 & $1250^3$ & $1250^3$ & $4.7\times 10^8$ & $8.8 \times 10^7$ & 2.96 & 375 \\
         & & TNG-300-3 & 302.6 & $625^3$ & $625^3$ & $3.8 \times 10^9$ & $7.0 \times 10^8$ & 6.05 & 757 \\
         & & \textbf{TNG-100} & \textbf{110.7} & $\mathbf{1820^3}$ & $\mathbf{1820^3}$ & $\mathbf{7.5 \times 10^6}$ & $\mathbf{1.4 \times 10^6}$ & \textbf{0.74} & \textbf{92.5} \\
         & & TNG-100-2 & 110.7 & $910^3$ & $910^3$ & $6.0 \times 10^7$ & $1.1 \times 10^7$ & 1.48 & 185 \\
         & & TNG-100-3 & 110.7 & $455^3$ & $455^3$ & $4.8 \times 10^8$ & $9.0 \times 10^7$ & 2.96 & 370 \\
         & & \textbf{TNG-50} & \textbf{51.7} & $\mathbf{2160^3}$ & $\mathbf{2160^3}$ & $\mathbf{4.5 \times 10^5}$ & $\mathbf{8.5 \times 10^4}$ & \textbf{0.29} & \textbf{36.3} \\
         & & TNG-50-2 & 51.7 & $1080^3$ & $1080^3$ & $3.6 \times 10^6$ & $6.8 \times 10^5$ & 0.58 &  72.5\\
         & & TNG-50-3 & 51.7 & $540^3$ & $540^3$ & $2.9 \times 10^7$ & $5.4 \times 10^6$ & 1.16 & 145\\
        \hline
        Dark & MillenniumTNG & \textbf{MTNG-740-Dark} & \textbf{740} & $\mathbf{4320^3}$ & --- & $\mathbf{1.95 \times 10^8}$ & --- & \textbf{3.7} & --- \\
        & & MTNG-185-Dark & 185 & $1080^3$ & --- & $1.95 \times 10^8$ & --- & 3.7 & --- \\
        & & MTNG-93-Dark & 93 & $540^3$ & --- & $1.95 \times 10^8$& --- & 3.7 & --- \\
         & IllustrisTNG & \textbf{TNG-300-Dark} & \textbf{302.6} & $\mathbf{2500^3}$ & --- & $\mathbf{5.9 \times 10^7}$ & --- & \textbf{1.48} & --- \\
         & & TNG-300-2-Dark & 302.6 & $1250^3$ & --- & $4.7 \times 10^8$ & --- & 1.48 & --- \\
        & & TNG-300-3-Dark & 302.6 & $625^3$ & --- & $3.8 \times 10^9$ & --- & 1.48 & --- \\
         & & \textbf{TNG-100-Dark} & \textbf{110.7} & $\mathbf{1820^3}$ & --- & $\mathbf{7.5 \times 10^6}$ & --- & \textbf{0.74} & --- \\
         & & TNG-100-2-Dark & 110.7 & $910^3$ & --- & $6.0 \times 10^7$ & --- & 0.74 & --- \\
         & & TNG-100-3-Dark & 110.7 & $455^3$ & --- & $4.8 \times 10^8$ & --- & 0.74 & --- \\
         & & \textbf{TNG-50-Dark} & \textbf{51.7} & $\mathbf{2160^3}$ & --- & $\mathbf{6.5 \times 10^5}$ & --- & \textbf{0.29} & --- \\
         & & TNG-50-2-Dark & 51.7 & $1080^3$ & --- & $5.2 \times 10^6$ & --- & 0.29 & --- \\
         & & TNG-50-3-Dark & 51.7 & $540^3$ & --- & $4.2 \times 10^7$ & --- & 0.29 & --- \\
        \hline
    \end{tabular}
\end{table*}

In this work, we extend the halo mass range by a further order of magnitude, hence probing objects ranging from dwarf galaxies to superclusters. This is made possible by combining the three IllustrisTNG realisations with the newer MillenniumTNG cosmological hydrodynamic simulations. The methodology and first results of the MillenniumTNG project have been presented in a series of works focusing on different subjects: galaxy clusters \citep{Pakmor_2023}, high-redshift galaxies \citep{Kannan_2023}, the halo model \citep{Hadzhiyska_2023a, Hadzhiyska_2023b}, galaxy clustering and halo statistics \citep{Bose_2023, Contreras_2023, Hernandez-Aguayo_2023}, the impact of baryons and massive neutrinos on weak lensing \citep{Ferlito_2023}, the intrinsic alignments of galaxies and haloes \citep{Delgado_2023}, and the refinement of semi-analytic models of galaxy formation \citep{Barrera_2023}.

The MillenniumTNG run follows essentially the same galaxy formation model as its predecessor IllustrisTNG (hereafter, the `TNG galaxy formation model'), but it comprises a much larger volume $(\sim 740 \, \rm Mpc)^3$, and a mass resolution of $\sim 3 \times 10^7 \, \rm M_{\odot}$ per baryonic mass element. We consider both the fully hydrodynamic runs and the DMO variants of all simulations. The combination of all runs enables us to study the impact of baryonic physics on the mass content and on the concentration-mass relationship of haloes in the mass range $10^{9.5} - 10^{15.5} \, \rm M_{\odot}$, and redshift interval $0<z<7$. To the best of our knowledge, this is the largest total halo mass and redshift range considered for this kind of study with cosmological hydrodynamic simulations. We test several empirical and physically motivated models for the concentration-mass relationship at different redshifts through a rigorous information criterion, and provide the best-fit parameters. The interested reader can thus readily use our tabulated results to model the DM density profiles with the TNG cosmology and galaxy formation model.

This manuscript is organised as follows. In Section~\ref{sec:simulation} we summarise the main characteristics of the IllustrisTNG and MillenniumTNG simulations. In Section~\ref{sec:results} we show how the halo mass of individual objects varies upon adding baryons in the simulations. We also show the concentration-mass relationship given by all runs considered, providing suitable analytic fitting formulae. In Section~\ref{sec:discussion}, we discuss the astrophysical implementation of our results, and compare them to previous similar work. We present our conclusions in Section~\ref{sec:conclusions}. 

Throughout this manuscript, unless otherwise stated, we indicate co-moving units with a `c' prefix (e.g., ckpc, cMpc, etc.).

\section{Simulations}
\label{sec:simulation}

In this work, we combine the publicly available suite of cosmological hydrodynamic simulations IllustrisTNG \citep{IllustrisTNG2018, TNGPublicDataRelease, TNG50_2019} with its successor MillenniumTNG. For all simulations, we consider both the full-physics hydrodynamic runs and their dark-matter-only (DMO) variants. Both IllustrisTNG and MillenniumTNG have been extensively described in the literature, therefore we will only briefly summarise the features that are most relevant for our work.

\begin{figure*}
    \centering
    \includegraphics[width=\textwidth]{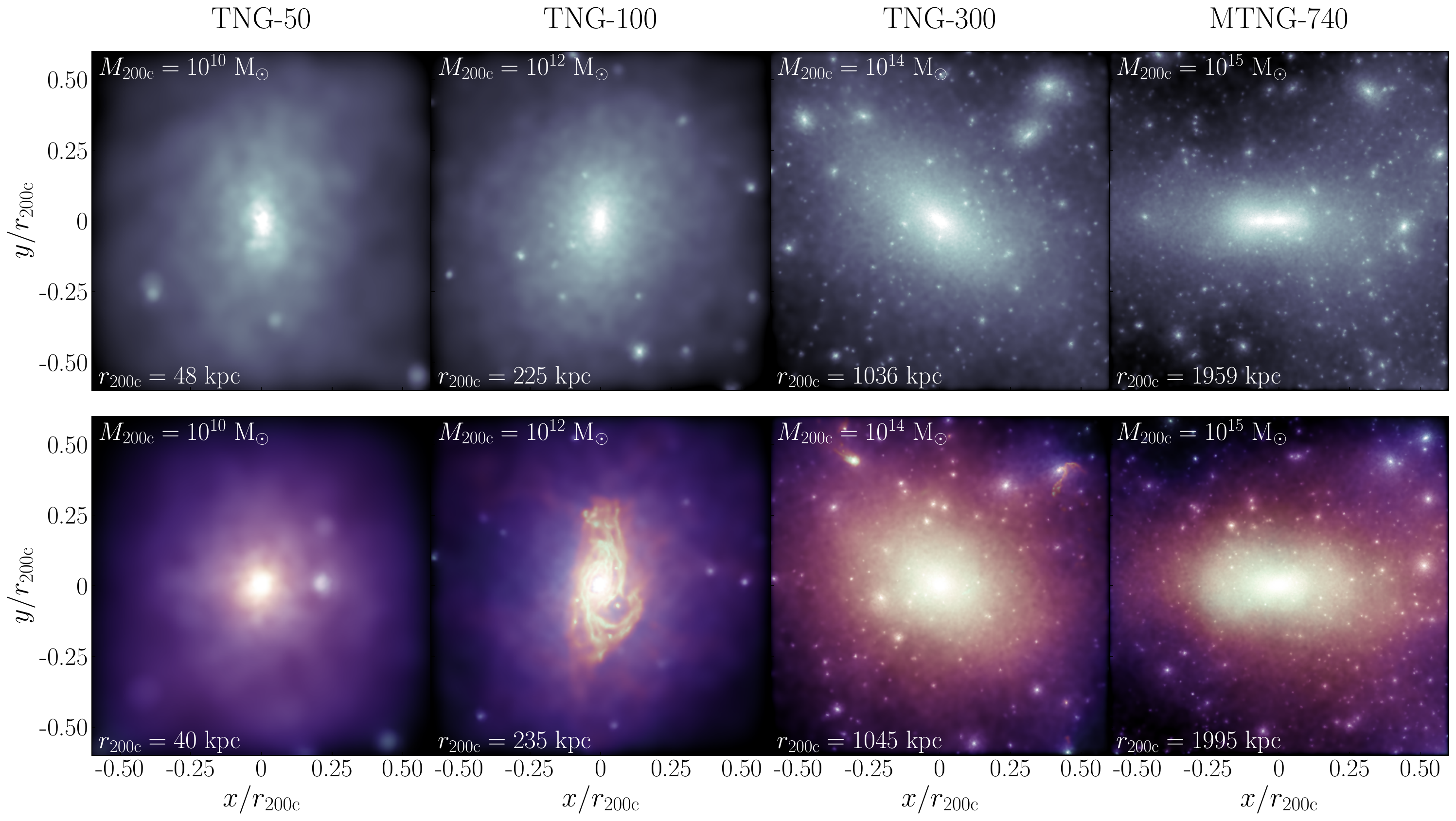}
    \caption{Projected density maps of haloes with different masses in the IllustrisTNG and MillenniumTNG hydrodynamic simulations (top panels), and of their counterparts in the DMO variants of the same simulations (see Section~\ref{sec:results} for details on the halo matching technique across the different runs). The upper panels show the dark matter density distribution, with lighter shades corresponding to regions with higher density. The lower panels overlay the projection of dark matter and gas density, represented with blue and purple-red-yellow colour maps, respectively. Also for the gas, lighter colours correspond to higher densities. This figure showcases the level of detail that can be achieved over an expansive dynamic range in total halo mass by combining simulations with different box sizes and mass resolutions (see Table~\ref{tab:simulations}).}
    \label{fig:haloes}
\end{figure*}

All simulations considered treat DM as self-gravitating Lagrangian particles within a fully Newtonian scheme with periodic boundary conditions, whereby the expansion of spacetime follows from the general-relativistic Friedman-Lemaitre-Robertson-Walker equations with null curvature. In the IllustrisTNG simulations, gravitational forces are calculated with a Tree-Particle-Mesh (Tree-PM) scheme \citep[following][]{Xu_1995, Bode_2000, Bagla_2002}, whereby the gravitational potential is divided in Fourier space into long-range and short-range components. The short-range interactions are computed through a hierarchical multipole expansion utilising an oct-tree structure \citep{Barnes_1986, Hernquist_1986}, which is adjusted by a short-range cut-off factor. Long-range interactions are derived from the potential achieved using the Fast Fourier Transformation mesh method, employing cloud-in-cell deposition to establish the mass density field on a uniform Cartesian grid. In the MillenniumTNG simulation, the same Tree-PM scheme is incorporated within an adjusted version of the \texttt{Gadget-4} code \citep{Gadget4}.

In all simulations, gas mass elements are hydrodynamically evolved on an unstructured Voronoi tessellation following the \texttt{Arepo} moving-mesh code \citep{Arepo}. The underlying physical framework is the IllustrisTNG galaxy formation model, which has been shown to effectively simulate a realistic galaxy population in a cosmological context \citep[see e.g.][]{Weinberger_2017, IllustrisTNG2018}. This model encompasses primordial and metal line cooling processes \citep{Vogelsberger_2013}, a sub-grid approach for the interstellar medium and star formation \citep{Springel_2003}, the recycling of mass and metals into the interstellar medium by AGB stars and Type Ia and II supernovae, a robust model for galactic outflows \citep{IllustrisTNG2018}, and a comprehensive mechanism for the growth of supermassive black holes and feedback from active galactic nuclei \citep[AGN;][]{Weinberger_2017}. 

The IllustrisTNG simulation employs a full magneto-hydrodynamical scheme, whereas magnetic fields were not followed in the  MillenniumTNG simulation due to memory constraints. Other adjustments were introduced to address minor shortcomings in the IllustrisTNG simulation that were discovered after it had been run \citep{TNGPublicDataRelease}, but the modifications are not expected to significantly affect the resulting galaxy formation history (see \citealt{Pakmor_2023} for details). Thus, the IllustrisTNG and MillenniumTNG galaxy formation schemes are effectively very similar, and that is why we simply refer to the `TNG galaxy formation model' in this manuscript.

All simulations identify structures and substructures on the fly. In the IllustrisTNG runs, this is accomplished via the friends-of-friends (FoF) and \texttt{SUBFIND} algorithms for haloes and subhaloes, respectively \citep{Springel_2005, Dolag_2009}. In the case of MillenniumTNG, subhaloes are identified with an adaptation of the more recent, \texttt{Gadget-4}-native \texttt{SUBFIND-HBT} algorithm into the \texttt{Arepo} moving-mesh code.

The Planck-2016 cosmology \citep{Planck16} is adopted in all simulations: $\Omega_{0}=0.3089$, $\Omega_{\rm b}=0.0486$, $\Omega_{\Lambda}=0.6911$, $\sigma_8 = 0.8159$, $n_{\rm s} = 0.9667$, and $h=0.6774$, with the usual definitions of the cosmological parameters. In the IllustrisTNG suite, initial conditions (ICs) at the starting redshift $z=127$ are generated via the \texttt{N-GENIC} code \citep{Springel_2005}. The ICs descend from the Zel'dovich approximation, applied to a particle distribution sampled from the linearly evolved matter power spectrum produced by the \texttt{CAMB} software \citep{CAMB_paper, CAMB_code}. For the MillenniumTNG runs, the ICs are produced following second-order Lagrangian perturbation theory with \textsc{Gadget4} at the initial redshift $z=63$. Following the fixed-and-paired variance suppression technique by \cite{Angulo_2016}, two realisations of the initial DM particle distribution are generated, each with the same mode amplitudes but opposite phases. The two realisations are designated as the `A' and `B' series (see \citealt{Hernandez-Aguayo_2023} for details). 

We summarise the main characteristics of the simulations utilised in this work in Table~\ref{tab:simulations}, together with the labels that we will use to refer to them in this manuscript. Throughout our analysis, we utilise all publicly available volumes of the IllustrisTNG, and the flagship MillenniumTNG run. We thus span a wide range of box sizes ($50 - 740 \, \rm cMpc$), which enables us to probe structures from dwarf galaxies to superclusters. For every run, we utilise the highest mass resolution available for the main analysis (highlighted in boldface type in Table~\ref{tab:simulations}), and reserve some lower-resolution variants for testing the robustness of our conclusions with appropriate convergence tests. We consider snapshots at redshift $z=0, \, 0.5, \, 1, \, 1.5, \, 2, \, 3, \, 4, \, 5, \, \mathrm{and} \, 7$. For the MillenniumTNG runs, we use only boxes from the `A' series.

\section{Results}
\label{sec:results}

In this section, we will present our findings on the impact of baryons on the total mass and on the dark matter density profiles within haloes. Throughout our analysis, we match haloes within the DMO runs of the MillenniumTNG simulation with their analogues in the corresponding hydrodynamic runs. This is possible because every DMO-hydrodynamic pair of simulations shares the same initial conditions for the DM particles. We can therefore extract the unique identifiers of the 16 most gravitationally bound particles within every halo of a given DMO run, and then find the halo sharing the largest fraction of those same particles in the corresponding hydrodynamic run. The shared fraction is determined by giving a higher weight to the particles that are more gravitationally bound, following the same method utilised to construct merger trees in \texttt{Gadget-4} \citep[see section 7.4 in][for further details]{Gadget4}. Thus, every halo in the hydrodynamic run hosting at least one of the particles in the halo originally considered in the DMO run is assigned a score based on the weighted number of particles shared. The halo with the highest score becomes the candidate to be matched with the original halo in the DMO run. In order to validate a link between two haloes, we repeat the procedure by swapping the hydrodynamic and DMO run, and retain only the bijective matches. This ensures that we do not inadvertently include spurious matches in our analysis that may arise from numerical artefacts connected to the halo finder, (see, e.g., the discussion in section~4.2 of \citealt{Sorini_2022}). In practice, less than 0.5\% of the haloes in the mass and redshift range considered in this work are discarded for not establishing a match. The matching technique described above is the same applied by \cite{TNGPublicDataRelease} to the IllustrisTNG simulations. We therefore use their publicly released catalogues of matched haloes when analysing properties of haloes drawn from the IllustrisTNG runs.

We show the results of the matching procedure for four haloes of different mass in Figure~\ref{fig:haloes} as an example. The upper panels show the 2D-projected dark matter mass density in the DMO variants of all simulations considered, as indicated in the top part of the figure. From left to right, we show haloes of increasing mass, as reported within the corresponding panel. Throughout the paper, we define the halo mass as $M_{\rm 200c}$, i.e. the total mass delimited by the spherically symmetric boundary $r_{\rm 200c}$, centred at the minimum of the gravitational potential, enclosing a matter mass density equal to 200 times the critical density of the Universe. The extent of each image is the same in units of $r_{\rm 200c}$, which we will adopt as the proxy for the virial radius in this work. We include the value of $r_{\rm 200c}$ within every panel for the reader's convenience.

The colour map in the upper panels of Figure~\ref{fig:haloes} shows the highest-density regions as white, and gradually switches to shades of blue in regions with less DM. Black areas are devoid of matter completely. The lower panels represent the matched counterparts of the haloes in the upper panels. In this case, the haloes are taken from the hydrodynamic simulations, thus they contain both DM and baryonic mass elements. We therefore overlay the 2D-projected DM and gas density maps. For the DM, we adopt the same colour coding as in the upper panels. The gas maps transition from bright yellow in the higher density regions to shades of red and eventually purple as the density diminishes. 

The gas density is broadly a smoother version of the underlying dark matter field, filling more uniformly the regions in between substructures. However, it also exhibits unique features, such as the spiral-shaped filaments that appear within $0.25 \, r_{\rm 200c}$ in the $10^{12}\, \rm M_{\odot}$ halo shown in Figure~\ref{fig:haloes}. These are presumably tracers of star-forming regions within the central galaxy of the halo. Feedback processes cause a diffuse distribution of gas, which contrasts with the more clumpy structure of DM. It is remarkable that such particulars are easily visible. Thanks to the different mass resolutions of the simulations considered, we are able to maintain a high level of detail for the matter density distribution over an expansive range of scales, ranging from dwarf galaxies ($M_{\rm 200c} \lesssim 10^{10}\, \rm M_{\odot}$) to superclusters ($M_{\rm 200c} \gtrsim 10^{15}\, \rm M_{\odot}$). This will ensure the robustness of our results, as we will demonstrate later in this section.

\subsection{Impact of baryons on the total halo mass}
\label{sec:mass_ratio}

\begin{figure}
    \centering
    \includegraphics[width=\columnwidth]{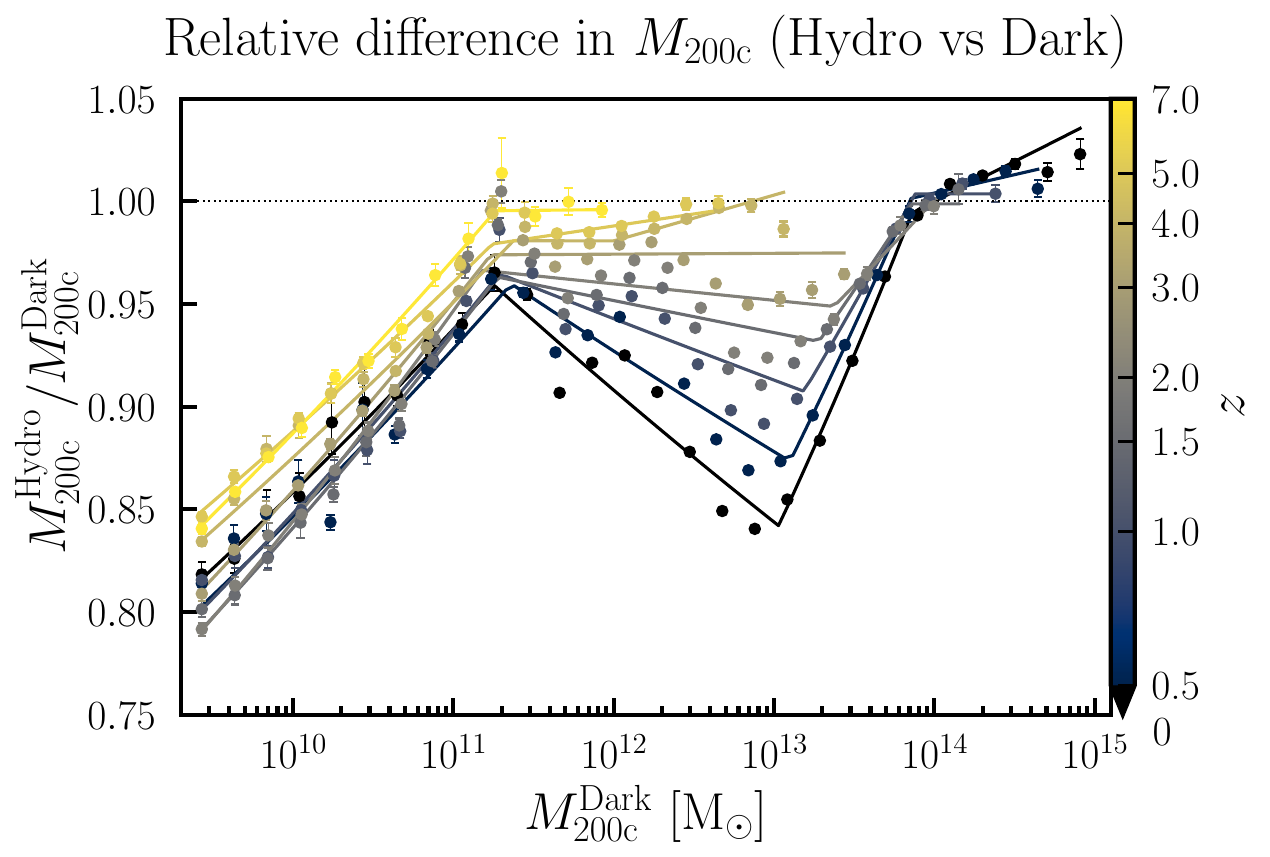}
    \caption{Ratio of the total mass (DM and baryons) of haloes in the hydrodynamic runs with respect to their matched counterparts in the dark-matter-only simulations, in the redshift range $0 \leq z \leq 7$. The circles correspond to the geometric mean of the ratio taken within equally extended logarithmic bins of the total halo mass in the DMO runs. The error bars indicate the statistical error on the geometric mean. The data points are colour coded according to the redshift of the snapshot to which they refer, as indicated in the colour bar. The thin solid lines plotted on top of the data sets represent the best-fit multiply broken power laws to the data (see Section~\ref{sec:mass_ratio} for details, and Table~\ref{tab:mass_ratio_params} for a list of the best-fit parameters at each redshift). The horizontal black dotted line marks a mass ratio of unity, to guide the eye. Two breaks of the power law are clearly identifiable around mass scales of $\sim 10^{11.3}\, \rm M_{\odot}$ and $\sim 10^{13} \, \rm M_{\odot}$. Above $M_{\rm 200c} \gtrsim 10^{14}\, \rm M_{\odot}$, the total halo masses in the hydrodynamic and dark-matter-only runs are equal within 1-2\%.
    }
    \label{fig:mass_ratio}
\end{figure}

To begin with, we focus on the impact of baryons on the total halo mass. To ensure that our results are converged, we restrict our analysis to haloes containing at least 3000 particles in all primary DMO runs (see Appendix~\ref{app:convergence} for details). For every snapshot, we bin all haloes according to their total mass, $M_{\rm 200c}$. The bins are constructed by taking the minimum and maximum halo masses in the snapshot considered, and dividing this range in logarithmic intervals with the same width of $0.2 \, \rm dex$. If the highest-mass bin contains fewer than 5 haloes, we merge it with the second-highest-mass bin, and reiterate the procedure until this condition is met. This ensures that the bin at the highest-mass end does not suffer from low-number statistics due to cosmic variance. We then match the haloes within each resulting mass bin with their analogues in the hydrodynamic run, following the matching technique described earlier. At this point, for every halo pair, we calculate the $M_{\rm 200c}$ (total mass) ratio between the hydrodynamic and DMO runs. 

We show the results of our analysis in Figure~\ref{fig:mass_ratio}. The $x$-axis represents the halo mass in the DMO run, and the $y$-axis the hydrodynamic-to-DMO mass ratio. The circles show the average ratio in each mass bin, estimated with the geometric rather than arithmetic mean. The advantage of such choice is that it can be straightforwardly inverted: the average DMO-to-hydrodynamic mass ratio is simply the inverse of the average hydrodynamic-to-DMO mass ratio. The points are colour coded according to the redshift of the snapshot to which they refer, as indicated in the colour bar. The error bars represent the statistical error on the geometric mean within each mass bin. Exploiting the fact that the geometric mean of a measurable quantity $X$ is the exponential of the arithmetic mean of $\ln (X)$, and applying the usual error propagation rules, the error on the geometric mean is given by:
\begin{equation}
        \sigma_{\langle \mathrm{MR} \rangle} = \langle \mathrm{MR} \rangle \frac{s[\ln(\mathrm{MR})]}{\sqrt{N}} \, ,
\end{equation}
where $\langle \mathrm{MR} \rangle$ is the geometric mean of the mass ratios $\rm MR$ of the $N$ halo pairs within the mass bin considered, and $s[\ln (\mathrm{MR})]$ is the sample standard deviation of the natural logarithm of the mass ratios.

We note that at high redshift ($z\geq 5$), the mass ratio is statistically within unity at a mass scale of $M_{\rm 200c} \gtrsim 10^{11} \, \rm M_{\odot}$. At lower masses, the ratio decreases, reaching $\sim 0.8$ at the lowest-mass end of $10^{9.3}\, \rm M_{\odot}$. This drop in the halo mass following the inclusion of baryons in a cosmological simulation has been observed in previous work \citep[e.g.][]{Sawala_2013}, and is connected to stellar feedback processes pushing gas elements well beyond the virial radius \citep{Sorini_2022, Ayromlou_2023}. As one considers haloes of higher mass, the momentum imparted by stellar-driven outflows \citep{IllustrisTNG2018} becomes progressively ineffective at overcoming the deeper gravitational potential well. This would explain the rise in the hydrodynamic-to-dark halo mass ratio up until $M_{\rm 200c} \approx 10^{11.3} \, \rm M_{\odot}$ \citep{Springel_2018}. At lower redshifts, the peak observed at this mass scale falls below unity by a only a few per-cent. 

Haloes above $M_{\rm 200c} \gtrsim 10^{11.3} \, \rm M_{\odot}$ exhibit a significant mass loss when baryons are included in the simulations. The decreasing trend continues until $M_{\rm 200c} \approx 10^{13}\, \rm M_{\odot}$. This is again consistent with previous numerical works \citep[e.g.][]{Illustris_V2014, Schaller_2015, Springel_2018}, as AGN-driven winds and jets are effective at displacing baryons from haloes, and preventing further gas accretion and star formation due to kinetic and thermal feedback \citep{Sorini_2022, Ayromlou_2023}.

\begin{figure}
    \centering
    \includegraphics[width=\columnwidth]{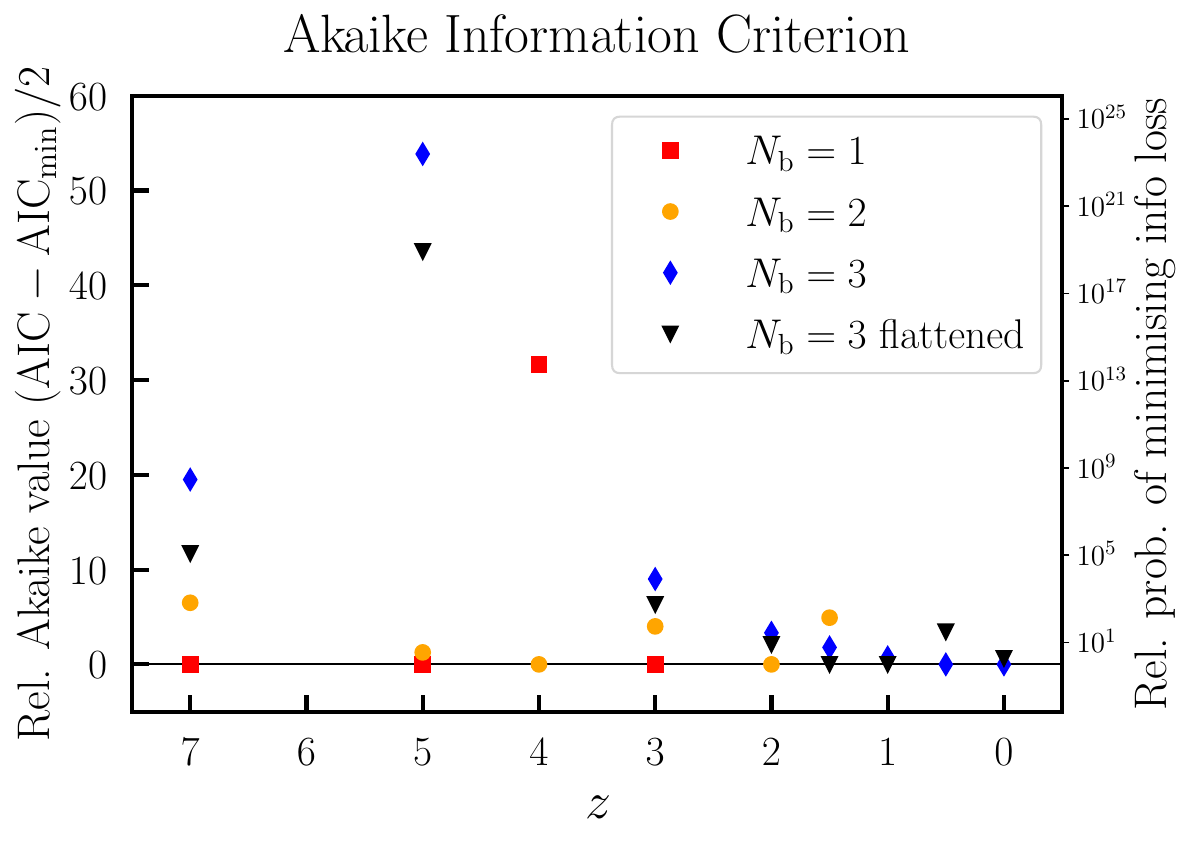}
    \caption{Relative Akaike's information criterion (AIC) value (equation~\ref{eq:akaike}) for each broken power-law model used to represent the numerical data obtained for the hydrodynamic-to-DMO mass ratio (Figure~\ref{fig:mass_ratio}), with respect to the best-fit model (see Section~\ref{sec:mass_ratio}). Every line refers to a different fitting function, as indicated in the legend. The ancillary $y$-axis reports the factor by which every model is \textit{less} likely to minimise the information loss, with respect to the best-fit model. At lower redshift ($z<2$), it is necessary to consider power laws with three breaks, while at higher redshifts simpler models are preferred by the AIC.}
    \label{fig:mass_ratio_akaike}
\end{figure}

Above $M_{\rm 200c} \approx 10^{13}\, \rm M_{\odot}$, the halo mass ratio increases again, saturating to unity (within a few per cent) at $M_{\rm 200c} \gtrsim 10^{14} \, \rm M_{\odot}$. Whereas AGN feedback is still active in these haloes, the gravitational potential is stronger due to the larger mass. Therefore, it becomes progressively harder for feedback processes to remove baryons from haloes, which approach the `closed-box' approximation \citep{Angelinelli_2022, Angelinelli_2023}. 

Our results at $z=0$ extend the analogous studies by based on the TNG-100 and TNG-300 simulations and their DMO counterparts \citep{Lovell_2018, Springel_2018}. We find the same qualitative trend for the hydrodynamic-to-DMO mass ratio, with transition points occurring at the same mass scales. Compared with the Illustris simulation, the TNG galaxy formation model is more efficient at decreasing $M_{\rm 200c}$ at the lower-mass end, while it exhibits a weaker imprint at the higher mass-end. This confirms the findings in \cite{Lovell_2018} and \cite{Springel_2018}, and reflects the differences in the underlying stellar and AGN feedback models, respectively, between the Illustris and IllustrisTNG/MillenniumTNG simulations. 

We additionally verified that if we consider the ratio between the DM mass enclosed within $r_{\rm 200c}$ of haloes in the hydrodynamic simulations and their DMO counterparts (properly corrected by a \smash{$1+f_{\rm b}$} factor), the resulting trend with $M^{\rm DMO}_{\rm 200c}$ is qualitatively similar to the one obtained in Figure~\ref{fig:mass_ratio} for the total halo mass ratio. However, when considering the DM component only, the maximum relative difference is reduced to $\sim 10\%$. This suggests that the shape of the total mass ratio as a function of $M^{\rm DMO}_{\rm 200c}$ and redshift is primarily driven by the presence of baryons, and is not merely a consequence of the redistribution of the DM component, which could alter $r_{\rm 200c}$, and hence $M_{\rm 200c}$. Thus, an analytical approximation of the numerical results would serve as a useful tool to imprint the effect of the TNG galaxy formation model on the total halo mass obtained from cheaper DMO simulations. We therefore provide empirical fitting formulae to our numerical results for the hydrodynamic-to-DMO mass ratio $\mathcal{R}$, as a function of the halo mass in the DMO runs, $M_{\rm 200c}^{\rm DMO}$. 

At any fixed redshift, we adopt a broken power law, defined as follows:
\begin{equation}
\label{eq:broken_pl}
    \mathcal{R}(M^{\rm DMO}_{\rm 200c}) = 
    C \left( \frac{M^{\rm DMO}_{\rm 200c}}{M_i} \right)^{\alpha_i}  \; \;
    \mathrm{for} \;\; M_{i-1} \leq M^{\rm DMO}_{\rm 200c} < M_{i} \, , 
\end{equation}
where $M_i$ refers to the mass scale corresponding to the $i$-th break of the power law. For a power law with $N_{\rm b}$ breaks, the index $i$ runs from $i=1$ to $N_{\rm b}+1$, so that $M_0$ and $M_{N_{\rm b}+1}$ refer, respectively, to the minimum and maximum halo mass in the entire range considered. With our indexing convention, it follows that $\alpha_i$ is the slope of the power law in the range $M_{i-1} \leq M^{\rm DMO}_{\rm 200c} < M_{i}$. The $C$ parameter simply regulates the normalisation of the power law and, by definition, corresponds to the value of the mass ratio at the break point $M^{\rm DMO}_{\rm 200c}=M_1$.

\begin{table}
\caption{Percentage of relaxed haloes in the DMO runs containing at least 5000 particles, and for which a bijective link with a halo in the corresponding hydrodynamic runs is established (see Section~\ref{sec:results}). From left to right, the columns show: the name of the simulation; the mass cut corresponding to 5000 DM particles, so that the haloes selected have a total mass $M_{\rm 200c}>M_{\rm cut}$; the redshifts of the snapshots considered.}

    \label{tab:relaxed}
    \centering
    \begin{tabular}{cccccc}
        \hline
         Simulation & $\log(M_{\rm cut}/\mathrm{M}_{\odot})$ & \multicolumn{4}{c}{$z$} \\
         &  & 0 & 1 & 3 & 7 \\
        \hline
         TNG-50-Dark & 9.5 & 66\% & 47\% & 17\% & 10\% \\
         TNG-100-Dark & 10.6 & 55\% & 34\% & 14\% & 12\% \\
         TNG-300-Dark & 11.5 & 43\% & 25\% & 14\% & 11\% \\
         MTNG-740-Dark & 12.0 & 40\% & 22\% & 12\% & 9.3\% \\
    \hline
    \end{tabular}
\end{table}

As discussed earlier in this section, the mass ratio exhibits between one and three mass scales causing a change in the slope, depending on redshift. To rigorously determine how many breaks to include in equation~\eqref{eq:broken_pl}, we first perform a minimum-$\chi^2$ fit to the numerical data at each redshift with a smoothed broken power law with one, two and three breaks. In the last case, we consider two variants, where the slope of the power law in the highest mass interval is either a free parameter or fixed to zero. This is motivated by the plateau that we observe at $M_{\rm 200c}  \gtrsim 10^{14} \, \rm M_{\odot}$ at $z<2$.

\begin{table*}
\caption{Parameters of the broken power laws fitting the hydrodynamic-to-DMO halo mass ratio in Figure~\ref{fig:mass_ratio}. See equation~\eqref{eq:broken_pl} and Section~\ref{sec:mass_ratio} for the definition of the parameters. If all haloes in the sample considered are below the mass threshold of $\sim 10^{13} \Msun$, then the hydrodynamic-to-DMO halo mass ratio is accurately described by the single-broken power law defined by the parameters $C$, $M_1$ and $\alpha_1$ reported in the table below. The only exception is $z=4$, where one would need haloes with $M_{\rm 200c}$ below $\sim 10^{12} \Msun$ in order to apply a single-broken power law.
}
    \label{tab:mass_ratio_params}
    \centering
    \begin{tabular}{ccccccccc}
        \hline
         $z$ & $C$ & $\log(M_1/\rmn{M}_{\odot})$ & $\log(M_2/\rmn{M}_{\odot})$ & $\log(M_3/\rmn{M}_{\odot})$ & $\alpha_1$ & $\alpha_2$ & $\alpha_3$ & $\alpha_4$\\
         \hline
        0 & $0.96 \pm 0.01$ & $11.26 \pm 0.01$ & $13.03 \pm 0.01$ & $13.86 \pm 0.01$ & $0.04 \pm 0.01$ & $-0.032 \pm 0.004$ & $0.086 \pm 0.008$ & $0.02 \pm 0.02$ \\
        0.5 &  $0.96 \pm 0.01$ & $11.4 \pm 0.3$ & $13.10\pm 0.01$ & $13.85 \pm 0.01$ & $0.04 \pm 0.02$ & $-0.024 \pm 0.003$ & $0.08 \pm 0.01$ & $0.01 \pm 0.02$ \\
        1.0* &  $0.965 \pm 0.009$ & $11.3 \pm 0.3$ & $13.19 \pm 0.01$ & $13.88 \pm 0.01$ & $0.04 \pm 0.02$ & $-0.014 \pm 0.003$ & $0.06 \pm 0.01$ & 0 \\
        1.5* &  $0.963 \pm 0.008$ &$ 11.3 \pm 0.3$ & $13.28 \pm 0.01$ & $13.85 \pm 0.01$ & $0.04 \pm 0.02$ & $-0.007 \pm 0.003$ & $0.05 \pm 0.02$ & 0\\
        2.0 & $ 0.966 \pm 0.007$ & $11.2 \pm 0.3$ & $13.37 \pm 0.01$ & --- & $0.05 \pm 0.02$ & $-0.004 \pm 0.003$ & $ 0.04 \pm 0.03$ & --- \\
        3.0 & $0.974 \pm 0.003$ & $11.2 \pm 0.01$ & --- & --- & $0.044 \pm 0.006$ & $ 0.000 \pm 0.002$ & --- & ---\\
        4.0 & $ 0.981 \pm 0.003$ & $11.38 \pm 0.09$ & $12.00 \pm 0.01$ & --- &$0.046 \pm 0.003$ &$ 0.000 \pm 0.003$ & $0.010 \pm 0.003$ & --- \\
        5.0 & $0.979 \pm 0.002$ & $11.24 \pm 0.06$ & --- & --- &  $ 0.034 \pm 0.001$ & $0.005 \pm 0.001$ & --- & --- \\
        7.0 & $ 0.995 \pm 0.007$ & $11.26 \pm 0.09$ & --- & --- &$ 0.040 \pm 0.01$ &$ 0.000 \pm 0.005$ & --- & ---\\
        \hline
    \end{tabular}
    \\
    \raggedright
    * {\footnotesize At these redshifts, the best-fit model is the flattened triple-broken power law. Thus, the parameter $\alpha_4$ is fixed to zero, as explained in Section~\ref{sec:mass_ratio}.}
\end{table*}

We then select the best fitting function by applying Akaike's information criterion (AIC; \citealt{Akaike_1974}). This criterion provides a hierarchy of the quality of different models in representing a given data set, by minimising the loss of information without overfitting. If $\widehat{\mathcal{L}}$ is the maximised value of the likelihood for a given model, and $k$ the number of free parameters, the corresponding AIC value is:
\begin{equation}
\label{eq:akaike}
    \mathrm{AIC} = 2 k - 2 \ln{\widehat{\mathcal{L}}} + \frac{2 k (k+1)}{n-k-1} \, ,
\end{equation}
where the last term introduces a correction for data samples of small size $n$. The best model is the one that minimises the AIC value. If the minimum value among the models considered is $\rmn{AIC}_{\rm min}$, then each model is a factor of $\smash{\exp[(\rmn{AIC}- \rmn{AIC}_{\rm min} )/2]}$ \emph{less} likely to minimise the information loss with respect to the best model. 

To calculate the maximum likelihood of each model given our data sets at any fixed redshift, we assume statistical independence of the data points. The expectation values are estimated by applying the fitting function to the mean halo mass in each bin, and the variances are given by the statistical error on the geometric mean of the hydrodynamic-to-DMO mass ratio. We then insert the maximum of the likelihood in equation~\eqref{eq:akaike}. The resulting AIC values relative to the best-fit model at each redshift are reported in Figure~\ref{fig:mass_ratio_akaike}. The horizontal black line marks the zero value, to guide the eye. A model lying on this line is the best model according to Akaike's information criterion. The ancillary $y$-axis shows how much less likely a given model is at minimising the loss of information, with respect to the best-fit model.

At higher redshifts, the simplest fitting function with one break only is preferred by the AIC. This is not surprising, because, down to redshift $z=5$, the halo mass range probed by the simulations encompasses only the smallest critical mass corresponding to a break in the hydrodynamic-to-DMO mass ratio  ($M_{\rm 200c}  \approx 10^{11} \, \rm M_{\odot}$). At redshift $2<z<4$, the second turnaround in the mass ratio corresponding to $M_{\rm 200c}  \approx 10^{13} \, \rm M_{\odot}$ starts becoming visible (Figure~\ref{fig:mass_ratio}), and the AIC favours a broken power law with two break points. However, at $z=3$, a power law with a single critical mass scale is marginally preferred. For $z<2$, the AIC selects a power law with three breaks, reflecting the higher complexity of the dependence of the mass ratio on $M_{\rm 200c}^{\rm DMO}$ over a wider mass range. We report the best-fit parameters of the fitting function preferred by the AIC in Table~\ref{tab:mass_ratio_params}.

Our fitting formulae are a useful analytical model that can be applied onto a DMO simulation to mimic the effect of the TNG galaxy formation model on the mass content of galactic haloes. We stress, however, that our model is purely empirical. Ideally, it would be preferable to fit a physically motivated function for the hydrodynamic-to-DMO mass ratio to the numerical data. This goes beyond the scope of the present manuscript, and will instead be the subject of a future investigation.

\subsection{Impact of baryons on density profiles}
\label{sec:dens_prof}

The analysis undertaken in the previous section, while informative, is agnostic to the detailed spatial distribution of DM within haloes. To gain further insight on this subject, we now analyse the DM density profiles as a function of halo mass and redshift in the hydrodynamic and DMO runs.

\begin{figure*}
    \centering
    \includegraphics[width=0.8\textwidth]{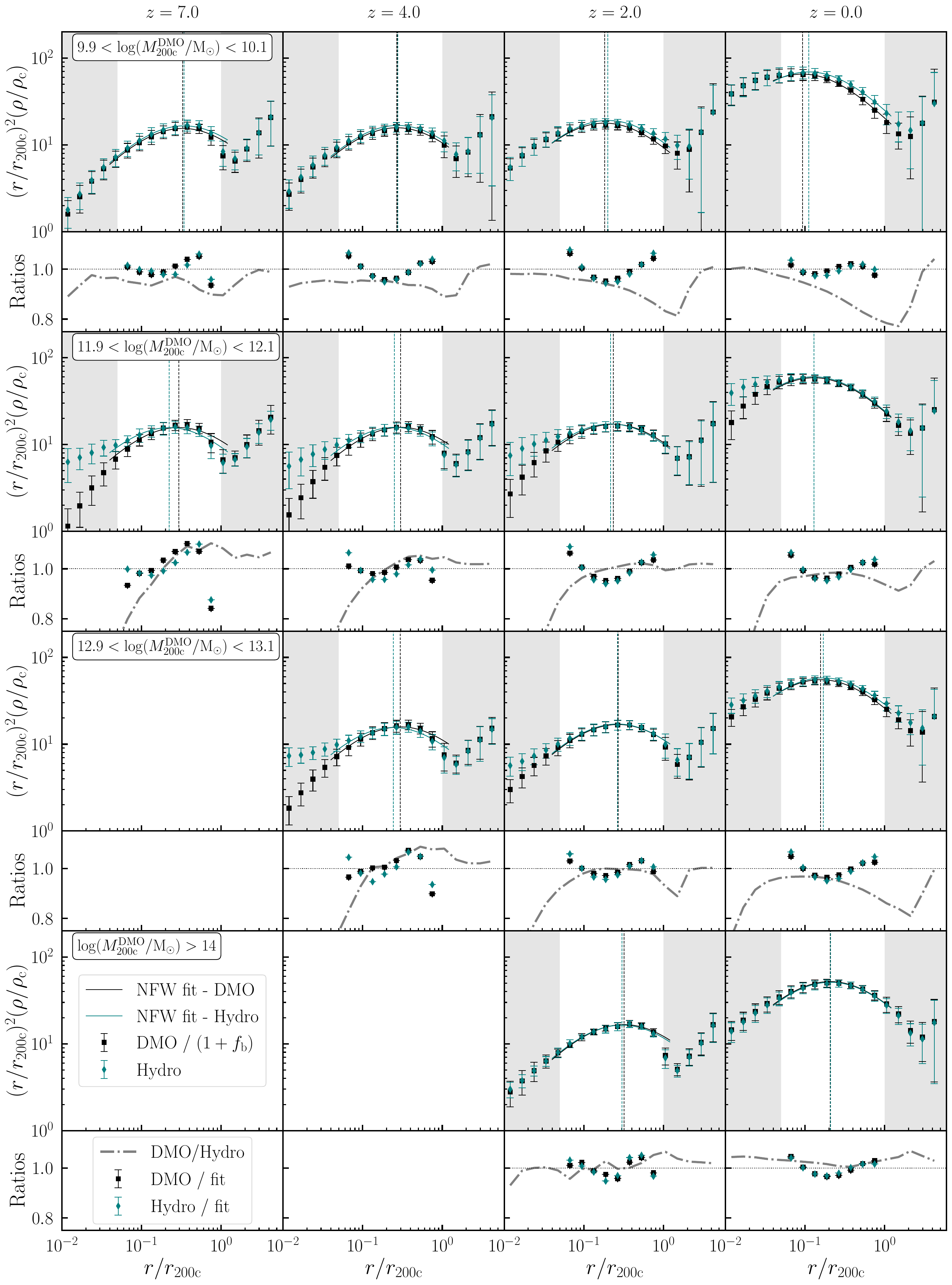}
    \caption{Redshift evolution of the dark matter density profiles within haloes of different mass within the DMO and hydrodynamic simulations considered in this study. Every row corresponds to a different redshift, as reported in each panel, and every column refers to a different halo mass bin, as indicated in the upper part of the figure. For a given redshift and halo mass bin, the upper panels show the average dark matter density profiles of haloes taken from the DMO runs (black squares), and their matched counterparts in the hydrodynamic runs (teal diamonds). The density profiles are normalised by the critical density of the Universe, and multiplied by the square of the halocentric distance, in units of $r_{\rm 200c}$. The error bars represent the $16^{\rm th}$ -- $84^{\rm th}$ percentile distribution of the density profile within each radial bin, across all haloes considered in the stack. To aid the readability of the figure, we omitted the lower error bar if the $16^{\rm th}$ percentile falls below the lower limit of the $y$-axis. The thin black and teal solid lines represent, respectively, the best-fit NFW profile \protect\citep{NFW} to the average dark matter density profile in the DMO and hydrodynamic runs. The vertical dashed lines mark the NFW scale radius resulting from the fit, following the same colour coding. Data points in the grey shaded area were excluded from the fit (see Section~\ref{sec:dens_prof} for details). The lower panels show the ratio between the profiles in the hydrodynamic and DMO simulations (grey dash-dotted lines), as well as the ratio between the profiles taken from the simulations and the best-fit NFW profiles (black and teal data points). Within the region where the fit was performed, the relative differences between simulation data and NFW fit remain within 10\%, regardless of halo mass and redshift. The NFW fitting functions accurately represent the density profiles in both the hydrodynamic and DMO runs.}
    \label{fig:dm_prof}
\end{figure*}

We first select all haloes containing at least 5000 particles in the DMO simulations. This selection criterion is more restrictive than the 3000 particles threshold that we adopted in Section~\ref{sec:mass_ratio} to analyse the hydrodynamic-to-DMO halo mass ratio. The reason is that we need to ensure that there are enough DM particles in any radial bin that we will be considering, in order to obtain a numerically reliable density profile. The choice of 5000 particles as the minimum requirement for haloes to be included in our analysis follows from previous similar works \citep{Schaller_2015}. 

From the resulting sample, we then select only relaxed haloes. We do this because we are primarily interested in the DM density profiles to study the effect of baryons on the concentration of DM haloes, and it is well known that haloes that recently underwent major mergers exhibit profiles that deviate more markedly from an NFW functional shape. Different criteria have been proposed in the literature to identify relaxed haloes, based on energetic considerations and the distribution of the DM halo mass across its substructures \citep[e.g.][]{Neto_2007}. \cite{Schaller_2015} verified that requiring the separation between the centre of mass of the halo and the centre of the minimum of the gravitational potential to be smaller than 7\% of its virial radius constitutes the most restrictive criterion for classifying a halo as `relaxed'. We therefore adhere to the same convention, adopting $r_{\rm 200 c}$ as a proxy for the virial radius. The fraction of relaxed haloes in the DMO simulations with at least 5000 particles, and for which a match with a halo in the corresponding hydrodynamic run has been established, is reported in Table~\ref{tab:relaxed} for a few representative redshifts. For a given simulation, the fraction of relaxed haloes increases at lower redshift, since, on average, more time has passed since the last major halo merger. At a fixed redshift, larger boxes contain a smaller fraction of relaxed haloes. This happens because the halo mass range probed by our larger-volume simulations is shifted towards higher masses, and more massive haloes tend to form later, hence having fewer time to reach dynamical relaxation.

For each relaxed halo, we define 20 radial bins as follows: the first bin spans the interval $0\leq r/r_{\rm 200c} < 0.01$, where $r$ denotes the 3D distance from the minimum of the gravitational potential of the halo; the remaining 19 bins span the range $0.01\leq r/r_{\rm 200c} < 5$ with equal width in logarithmic space. The DM density within each radial bin is then straightforwardly computed as the ratio of the total mass of all DM particles falling within said bin (not only those belonging to the FoF group), and the volume enclosed between the spherical shells defined by the boundaries of the bin. We then compute the DM density profiles with the same technique for the haloes in the hydrodynamic runs that match the relaxed haloes in the DMO runs as described in the beginning of Section~\ref{sec:results}.

We take a first look at the evolution of the DM density profiles across redshift and as a function of the total halo mass in Figure~\ref{fig:dm_prof}. For a given snapshot, we first combine the density profiles extracted from the DMO variants of all simulations, and organise them in mass bins, as annotated in the figure. We then select the density profiles from the matched haloes within the hydrodynamic runs. Since the matching technique is based on the unique IDs of the most tightly bound DM particles, the total mass of some of the matched haloes may in principle fall outside the boundaries of the DMO mass bin originally considered. However, in Figure~\ref{fig:dm_prof} we still associate them with the same mass bin defined for the DMO run, to guarantee a fair comparison between the hydrodynamic and DMO simulations. For all haloes at the selected redshift and DMO halo mass range, we then take the arithmetic mean of the co-moving DM density in each radial bin. The resulting average co-moving DM density profiles are represented in Figure~\ref{fig:dm_prof} (bigger panels) with black squares and teal diamonds for the DMO and hydrodynamic simulations, respectively. The density profiles in the DMO run are corrected by a $(1+f_{\rm b})$ factor, where $f_{\rm b}$ is the cosmic baryon fraction, for a fair comparison with the results of the hydrodynamic simulations. The error bars show the $16^{\rm th}-84^{\rm th}$ percentile of the DM density distribution across all haloes in any given radial bin. We follow the established practice of normalising the density profiles by the critical density of the Universe $\rho_{\rm c}$, and multiplying them by the square of the halocentric radial distance in units of $r_{\rm 200c}$ \citep[see, e.g.][]{Schaller_2015}; this makes it easier to infer the halo concentration, as we will explain later. However, for the sake of simplicity, we will refer to both $\rho(r)$ and $r^2 \rho(r)$, properly normalised, as `density profile' throughout this manuscript. 

In the smaller panels of Figure~\ref{fig:dm_prof}, we plot the ratios between the DM density profile given by the DMO runs and the hydrodynamic simulations (dot-dashed grey line). We see that the relative difference is generally contained within $10\%$. At $z=0$, the discrepancy can reach $\sim 20\%$ around the virial radius. For $M_{\rm 200c} \approx 10^{12}\, \rm M_{\odot}$ and $M \approx 10^{13} \, \rm M_{\odot}$, the density profiles in the hydrodynamic runs deviate by more than 20\% from their DMO counterparts in the innermost regions of the haloes ($r \lesssim 0.05 \, r_{\rm 200c}$), even at higher redshift. However, such differences are primarily driven by numerical artefacts rather than physical reasons. It is well known that the finite mass resolution of N-body simulations can introduce spurious effects on the density profiles in the central regions of haloes. With a suite of simulations of individual Milky-Way-mass haloes, \cite{Power_2003} found that numerical convergence is achieved at radii that contain enough particles such that the local two-body relaxation timescale is on par with or longer than a Hubble time. This condition defines the so-called `convergence radius', which can be more easily estimated from the box size and number of DM particles in a simulation thanks to the formula introduced by \cite{Ludlow_2019}. Using their equations 17-18, we verified that in our sample of haloes the convergence radius is of the order of 5\% of the virial radius. For this reason, we exclude data points in the region $r < 0.05 \,  r_{\rm 200c}$ from any further analysis. Since we focus on the distribution of DM within the halo only, we ignore all particles outside the virial radius of the halo.

We thus fit every mean density profile shown in Figure~\ref{fig:dm_prof} with an NFW profile, over the range $0.05 < r/r_{\rm 200c} < 1$. The best-fit parameters in equation~\eqref{eq:NFW} are determined via $\chi^2$ minimisation. For the concentration parameter, we adopt the definition $c_{\rm 200c} = r_{\rm 200c}/r_{\rm s}$, where $r_{\rm s}$ is the scale radius of the NFW profile. The scale radius in the DMO and hydrodynamic simulations, normalised by the mean $r_{\rm 200c}$ in the halo mass bin considered, is shown in Figure~\ref{fig:dm_prof} with vertical dotted black and teal line, respectively. The corresponding NFW fits to the density profiles are plotted with the solid lines following the same colour coding. It can be seen that the scale radius corresponds to the maximum of $r^2 \rho(r)$. This follows directly from the definition of the $y$-axis, and represents the main advantage of plotting $r^2 \rho(r)$ rather than the bare density profile.

We plot the ratio of the numerical density profiles with respect to the best-fit NFW profile in the smaller panels of Figure~\ref{fig:dm_prof}. The relative differences are always within $10\%$, meaning that the NFW profile describes the data within this level of accuracy. For any fixed halo mass, the normalised scale radius moves to larger values at lower redshift. This happens more rapidly for haloes in the hydrodynamic simulations. Such haloes are more concentrated than their DMO counterparts at high redshift for $M_{\rm 200c} < 10^{14} \, \rm M_{\odot}$. At lower redshift, the difference in concentration becomes smaller, and haloes in the smallest mass bin are less concentrated in the hydrodynamic runs than in the DMO variants at $z=0$. Instead, haloes from both type of simulations appear to have similar concentrations in the highest mass bin.

\begin{figure*}
    \centering
    \includegraphics[width=\textwidth]{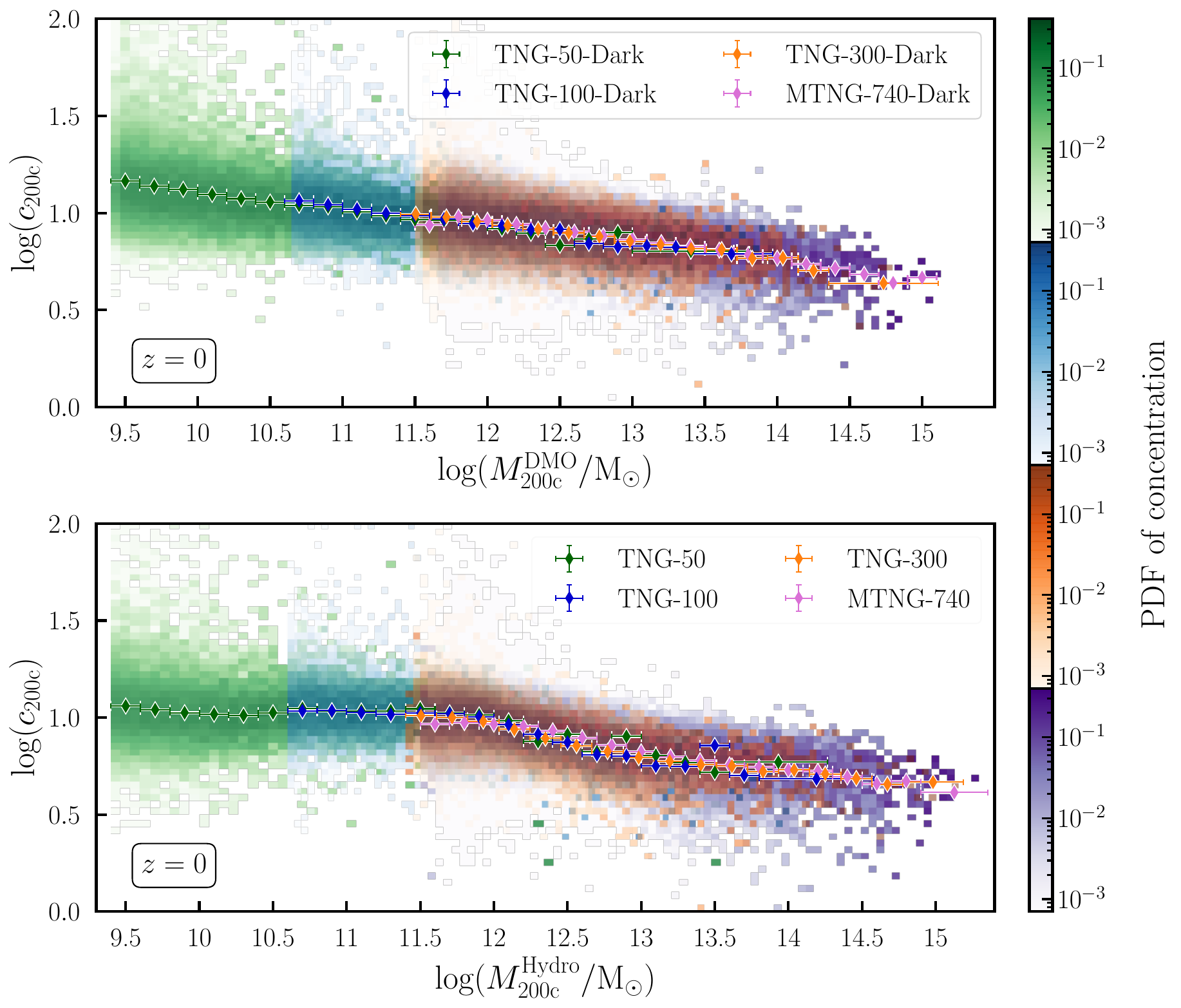}
    \caption{Concentration-mass relationship at $z=0$ for relaxed haloes in the DMO and hydrodynamic simulations (top and bottom panels, respectively). The overlapping colour maps represent the probability density function of the concentration of haloes within a fixed total halo mass bin. The data points represent the geometric mean of the halo mass in each bin. Each colour refers to haloes taken from a different simulation; from the smallest to the largest volumes, they are represented with shades of green, blue, orange and purple, respectively. The data points refer to the concentration of the mean density profile of the haloes within the mass bin delimited by the horizontal error bars. Points with different colours refer to different simulations, as indicated in the legend. The extended mass range in our work shows that the inclusion of baryons suppresses the concentration of haloes for $M_{\rm 200c} \lesssim 10^{11.5}\, \rm M_{\odot}$. The concentration-mass relationship in the hydrodynamic runs and, to a lesser extent, in the DMO runs, deviate from a pure power law.
    }
    \label{fig:cM_rel}
\end{figure*}

The density increase that we observe beyond $r_{\rm 200c}$ in all panels appears because the density profiles in Figure~\ref{fig:dm_prof} are computed from all particles within a given halocentric distance, and not only those included in the FoF group. The upturn at large radii is therefore induced by the two-halo term, representing the contribution due to matter external to haloes. However, in the remainder of this work we only focus on the impact of baryons on the internal structure of DM haloes, i.e., within $r_{\rm 200c}$, focusing on the dependence of the concentration on halo mass and redshift. A rigorous analysis in this sense will be the subject of the next section.

\subsection{Impact of baryons on the concentration-mass relationship}
\label{sec:conc-mass}

\subsubsection{Present-day concentration-mass relationship}
\label{sec:conc_z0}

We now analyse the concentration-mass relationship in the DMO and hydrodynamical simulations, for all snapshots considered. This will provide us with useful insights on the impact of baryons on the concentration of DM haloes.

To begin with, we focus on the DMO runs at $z=0$. For each simulation, we select well-resolved, relaxed haloes as explained in Section~\ref{sec:dens_prof}. We then construct mass bins with equal logarithmic width of $0.2 \, \rm dex$, starting from the minimum mass in the sample. If the highest-mass bin contains fewer than 5 haloes, we merge it with the previous bin. We then construct the DM density profile of all haloes falling in each bin, and compute the average profile, exactly as we did for Figure~\ref{fig:dm_prof} (see Section~\ref{sec:dens_prof}). 

In the upper panel of Figure~\ref{fig:cM_rel} we plot the concentration-mass relationship of the mean density profiles given by every DMO simulation with data points of different colours. The horizontal error bars represent the width of the mass bins. We also show the 2D histograms of the concentration-mass relationship resulting from fitting the density profiles of individual haloes with an NFW function. The histograms are represented with maps following the same colour coding as the data points, and are overlaid to simultaneously display the spread around the average concentration-mass relationship in the different simulations.

We then match all haloes from each simulation to their hydrodynamic counterparts, as explained in Section~\ref{sec:results}, and bin the haloes according to their total mass in the hydrodynamic run, following the same procedure adopted for the DMO runs. The concentration-mass relationship for the hydrodynamic simulations is then obtained with the same analysis described earlier for the DMO runs, and the results are shown in the lower panel of Figure~\ref{fig:cM_rel}, following the same colour coding as in the upper panel. We reiterate that, following the method just described, only the \textit{mean} density profile of the haloes within a given bin is used to estimate the concentration at the corresponding halo mass. We fit the profiles of individual haloes only to assess the spread around the concentration-mass relationship, as shown in Figure~\ref{fig:cM_rel}.

The data from different simulations are consistent with one another in the regions of the plot where they overlap. This suggests that the results are robust under different box sizes and mass resolutions, for all runs considered. We will further quantify the degree of numerical convergence across the different runs in Section~\ref{sec:cM_model} and in the Appendix~\ref{app:convergence}. But Figure~\ref{fig:cM_rel} already indicates that we can trust the concentration-mass relationship over a halo mass range of six orders of magnitude.

\begin{figure*}
    \centering
    \includegraphics[width=0.99\textwidth]{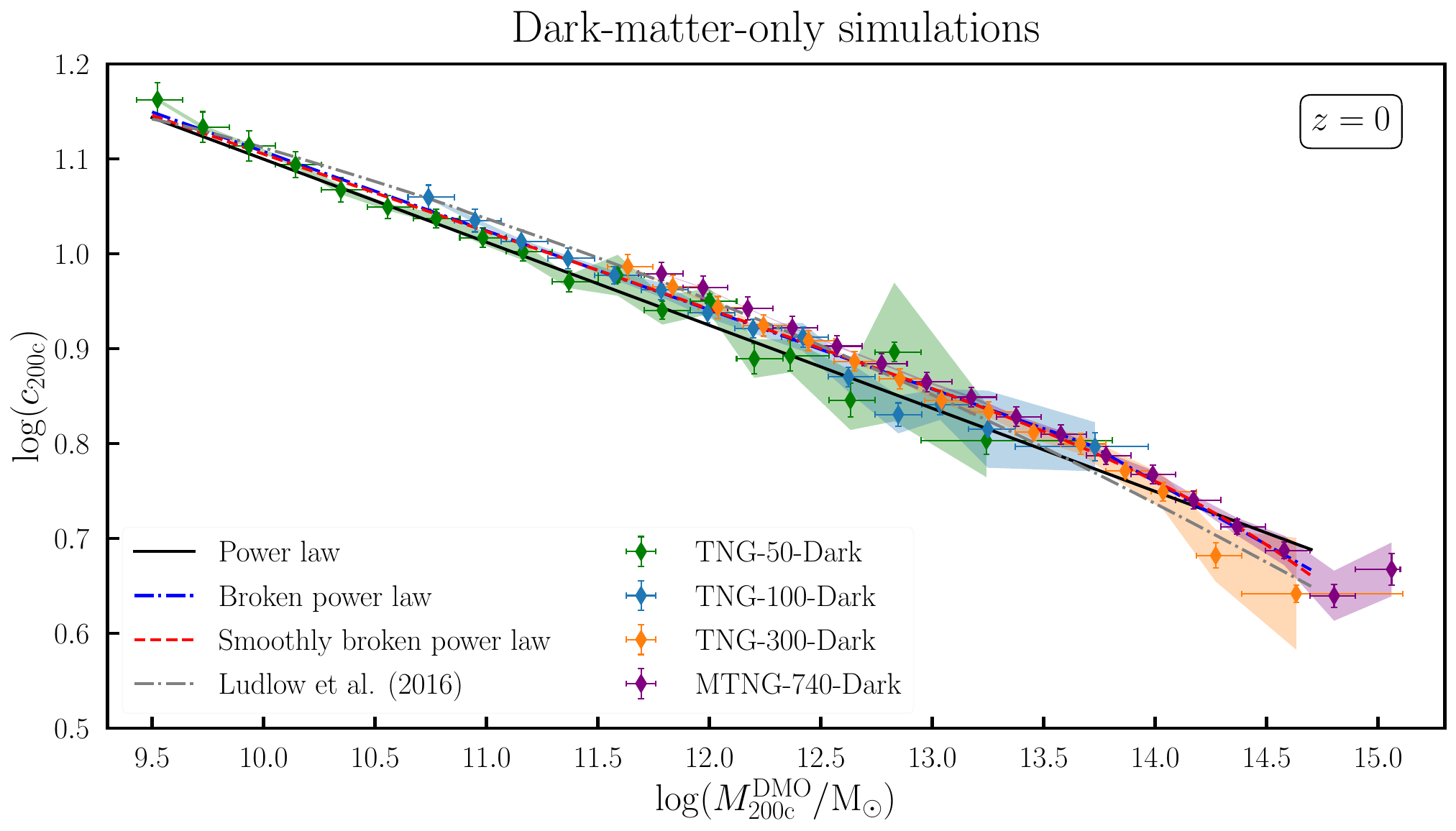}
    \hfill
    \includegraphics[width=0.99\textwidth]
    {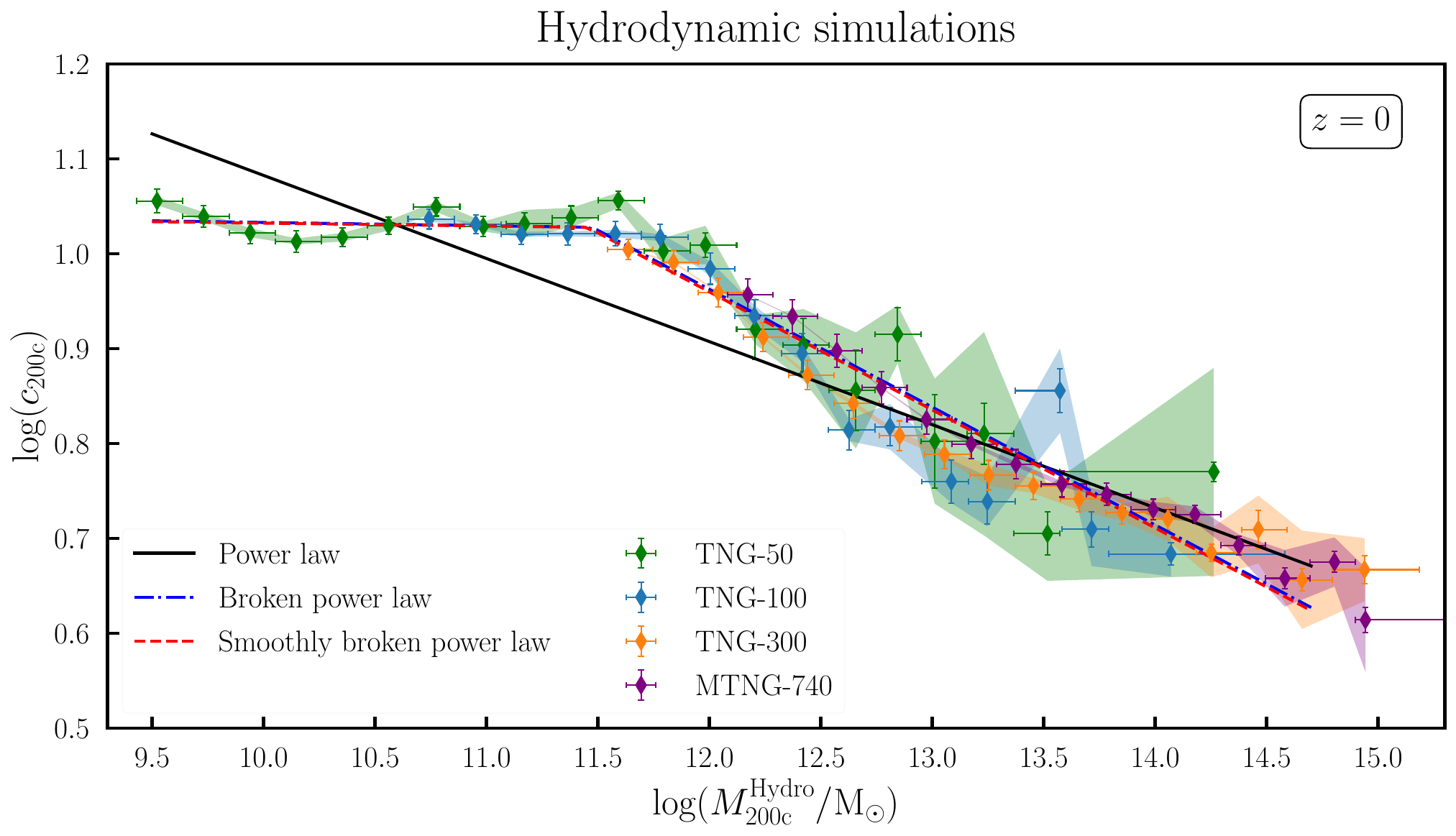}
    \caption{Concentration-mass relationship at $z=0$ for relaxed haloes in the DMO and hydrodynamic simulations (top and bottom panels, respectively). Data points represent the concentration of stacked density profiles within halo mas bins delimited with the horizontal error bars. The data point is plotted at the median halo mass within each bin. The vertical error bars represent the statistical error on the concentration deriving from the NFW fit. The shaded areas following the same colour coding as the data points show the statistical error due to cosmic variance or bootstrap re-sampling of the haloes, whichever is the largest (see Section~\ref{sec:cM_model} for details). Different best-fit models to the data are plotted with different colours and line styles, as indicated in the legend. The first-principles model of \protect\cite{Ludlow_2016} provides a good match to the DMO concentration-mass relationship, but broken power-law fits are best at representing the data in both the DMO and hydrodynamic simulations. A pure power law is still an acceptable fit for the DMO runs, but fails at reproducing the data once baryons are included. 
    }
    \label{fig:cM_fit}
\end{figure*}

The most obvious trend is that the concentration-mass relationship in the DMO simulations is monotonically decreasing with mass. This is a feature that has been repeatedly observed in N-body simulations \citep[e.g.][]{NFW, Dutton_2014, Schaller_2015, Beltz-Mohrmann_2021}. The spread around the average relationship is larger at the lower-mass end. This is not surprising either, since lower-mass haloes typically reside in more diverse environments than higher mass haloes, which leads to a spread in the formation time \citep{Harker_2006}. Furthermore, due to purely statistical reasons, lower-mass haloes are more likely to populate the tails of the distribution of the concentration, since they are present in greater abundance. Indeed, even at the higher-mass end, the range of observed concentration values is larger once we increase the box size, due to the larger number of massive haloes. This can clearly be seen from the results of the TNG-300-Dark and MTNG-740-Dark simulations at $M_{\rm 200c}\approx 10^{14} \, \rm M_{\odot}$. 

Similar considerations regarding the average trend and scatter apply to the hydrodynamic simulations as well. However, the slope of the relationship varies more strongly with mass. For $M_{\rm 200c}\lesssim 10^{11.5} \, \rm M_{\odot}$, the average concentration-mass relationship is almost flat, and certainly less steep than in the DMO case. Above such a mass threshold, the concentration declines more rapidly with mass, until $M_{\rm 200c} \approx 10^{13} - 10^{13.5} \, \rm M_{\odot}$. For higher masses, the slope becomes once again more gentle. As expected, the aforementioned mass scales roughly correspond to the breaks in the hydrodynamic-to-DMO halo mass ratio (Figure~\ref{fig:mass_ratio}). This is consistent with the fact that the impact of baryons on the mass content of haloes and their concentration are interconnected. 

In the remainder of the section, we will focus on the modelling of the concentration-mass relationship, and we will discuss the possible physical origins of any departure between the hydrodynamic and DMO results in Section~\ref{sec:discussion}.

\subsubsection{Modelling the concentration-mass relationship}
\label{sec:cM_model}

The concentration-mass relationship is such a crucial quantity in the context of cosmological structure and galaxy formation that numerous modelling attempts have appeared in the literature. These include empirical or first-principles analytical models, as well as semi-analytical or fully numerical methods \citep[e.g.][]{Bullock_2001, Gao_2008, Zhao_2009, Prada_2012, Dutton_2014, Ludlow_2013, Ludlow_2014, Beltz-Mohrmann_2021, Shao_2023, Shao_2024} In this section, we will therefore assess the success of different functional shapes at capturing the behaviour of the average concentration-mass relationship in our simulations.

Before testing any model for the concentration-mass relationship, we need to assess the statistical error in our data. We do so using three different methods. To begin with, the $\chi^2$-minimisation method for determining the best NFW fit to the mean density profile within each mass bin provides us with an estimate of both the mean and standard deviation of the concentration. However, such a standard deviation might be underestimating the statistical error on the average concentration in the mass bins that contain fewer haloes, where the PDF of the concentration may be deviating more strongly from a Gaussian distribution. As a second estimate, we therefore compute the standard deviation of the concentration by bootstrapping the density profiles in each mass bin. We consider 1000 samples with size equal to the number of haloes, allowing for repetitions; such procedure was verified to guarantee an accurate estimate of the sample variance in a previous similar work \citep{Brown_2022}. Finally, as our third estimate, we compute the cosmic variance on the concentration parameter in each mass bin by jackknife resampling of the haloes in a given mass bin upon dividing the simulation box in eight octants.

In Figure~\ref{fig:cM_fit}, we report the average concentration-mass relationships already shown in Figure~\ref{fig:cM_rel}. The vertical error bars represent the statistical error on the concentration arising from the NFW fit, i.e., following the first method described above. The shaded areas show the maximum between the bootstrap and cosmic variance errors, which we nevertheless verified to be of the same order of magnitude for all mass bins. As expected, the error from the NFW fit underestimates the spread of the average concentration in the mass bins with fewer haloes, i.e. at the higher-mass end. On the contrary, the error from the fit dominates at the lower-mass end. We make the conservative choice of considering the statistical error on the average concentration in each mass bin to be the maximum amongst the error from the fit, the bootstrap error, and cosmic variance.

At this point, we are able to determine the best-fit parameters of different models for the concentration-mass relationship via $\chi^2$ minimisation. The first model that we consider is a power law, which serves as a useful baseline due to its mathematical simplicity and widespread usage in the literature \citep{Dutton_2014, Schaller_2015, Ragagnin_2019, Ragagnin_2021, Beltz-Mohrmann_2021}. While this model appears to adequately represent the data in the DMO case, it is clearly oversimplified for the hydrodynamic simulations. We therefore introduce a broken power law, which yields an excellent agreement with the DMO data, and also allows us to capture the flattening of the concentration-mass relationship for $M_{\rm 200c} \lesssim 10^{11.5} \, \rm M_{\odot}$ observed in the hydrodynamic runs. We also consider a smooth variant of this model, which provides a continuous transition between the two power-law regimes across the mass scale, offering a more realistic representation of the gradual changes observed in the simulations.

\begin{table}
    \centering
    \textbf{Dark-matter-only simulations}
    \begin{tabular}{ccc}
        \hline
        Model &  $\Delta \rm AIC$ & $p_{\rm AIC}$ \\
        \hline
         Power law & $-16.6$ & $2.52 \times 10^{-4}$ \\
         Broken power law & $0$ & --- \\
         Smoothly broken power law & $-1.63$ & 0.443 \\
         \cite{Ludlow_2016} & $-62.5$ & $2.67 \times 10^{-14}$ \\
         \hline
    \end{tabular}
    \centering
    \vspace{5mm}
    \textbf{\\ Hydrodynamic simulations}
    \begin{tabular}{ccc}
        \hline
       Model &  $\Delta \rm AIC$ & $p_{\rm AIC}$ \\
       \hline
         Power law & $-483$ & $ 1.57 \times 10^{-105}$  \\
         Broken power law & $1$ & ---  \\
         Smoothly broken power law & $-3.88$ & 0.144  \\
         \hline
    \end{tabular}
    \caption{Results of the AIC test on the models considered as possible fits to the concentration-mass relationship at $z=0$. The columns show, starting from the left: the name of the model as mentioned in the text; the relative AIC value with respect to the best model according to the criterion; the corresponding probability of minimising the loss of information, with respect to the best model; the reduced $\chi^2$.}
    \label{tab:AIC}
\end{table}

\begin{figure}
    \centering
    \includegraphics[width=\columnwidth]{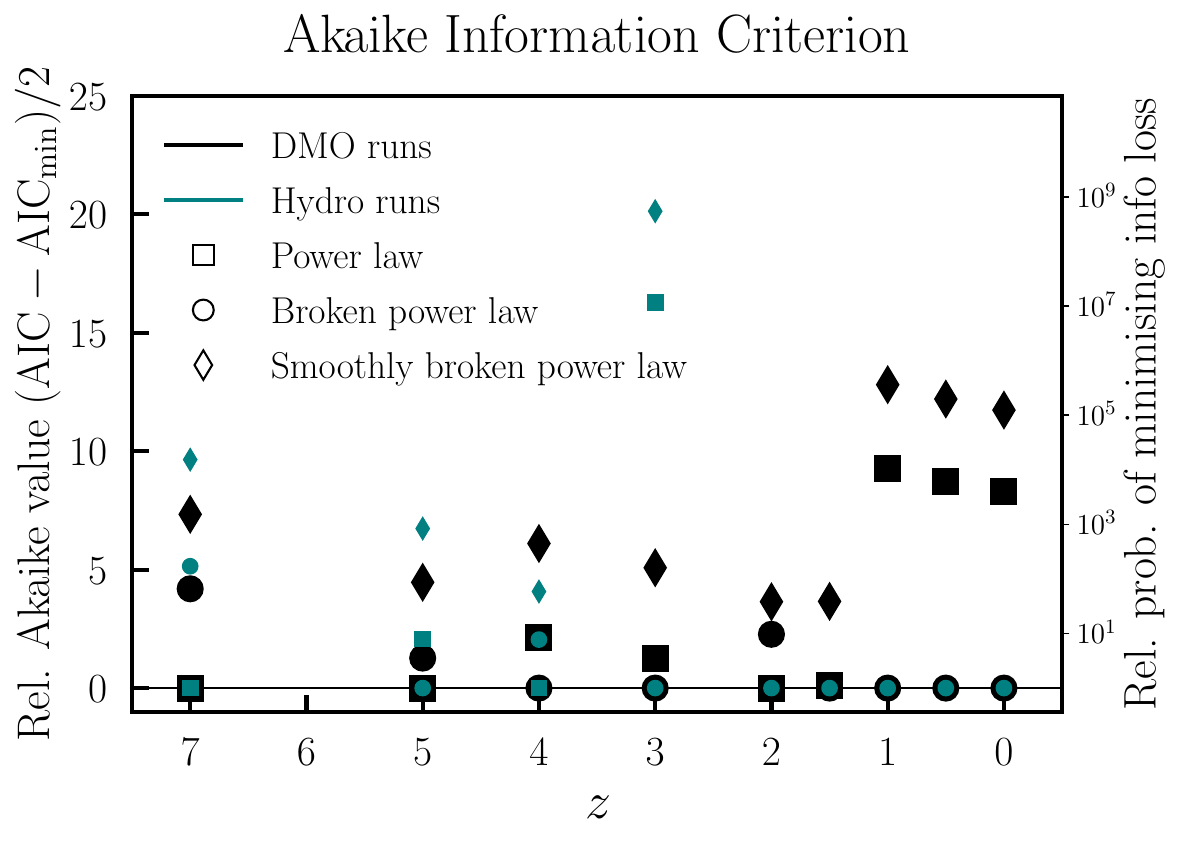}
    \caption{Relative Akaike's information criterion (AIC) value (equation~\ref{eq:akaike}) for each model (equation~\ref{eq:broken_pl}) used to represent the numerical data obtained for the concentration-mass relationship, with respect to the best-fit model, as a function of redshift. Every marker style refers to a different fitting function, as indicated in the legend. Black and teal points represent the results from the DMO and hydrodynamic runs, respectively. The ancillary $y$-axis reports the factor by which every model is \textit{less} likely to minimise the information loss, with respect to the best-fit model. The broken power law is the preferred model for most redshifts. The best-fit parameters for each redshift are reported in Table~\ref{tab:fit_params}. }
    \label{fig:cM_akaike}
\end{figure}

\begin{table*}
    \centering
    \caption{Best-fit parameters of the model for the concentration-mass relationship in the DMO and hydrodynamic simulations, as determined by the Akaike information criterion (see Figure~\ref{fig:cM_akaike}). The models considered are defined in equation~\eqref{eq:conc_bpl}. The parameter $A$ represents the normalisation of the relationship at the halo mass scale $M_{\rm ref}$; for a pure power law, we fixed such pivot scale to $10^{12}\, h^{-1} \, \rm M_{\odot}$, therefore it is not a free parameter of the model. The parameters $\alpha$ and $\beta$ represent the slopes of the relationship for $M\leq M_{\rm ref}$ and $M>M_{\rm ref}$, respectively. In the case of a pure power law, there is no $\beta$ parameter.
    }
    \label{tab:fit_params}
    \begin{tabular}{cccccccccc}
     \hline
        \multirow{2}{*}{\vspace{-0.25ex} $z$}  & \multicolumn{4}{|c|}{\bf Dark-matter-only simulations \vspace{0.25ex}} & & \multicolumn{4}{|c|}{\bf Hydrodynamic simulations \vspace{0.25ex}} \\
        \cline{2-5} \cline{7-10} \vspace{-2ex}\\
        
          & $A$ & $\log(M_{\rm ref}/\mathrm{M}_{\odot})$ & $\alpha$ & $\beta$ & & $A$ & $\log(M_{\rm ref}/\mathrm{M}_{\odot})$ & $\alpha$ & $\beta$\\
         \hline
         $0$ & $6.3 \pm 0.2$ & $13.71 \pm 0.02$  & $-0.083 \pm 0.002$ & $-0.13 \pm 0.01$ & & $10.7 \pm 0.2$ & $11.474 \pm 0.009$ & $-0.003 \pm 0.009$ & $-0.124 \pm 0.004$ \\
        $0.5$ & $6.9 \pm 0.3$ & $12.52 \pm 0.04$ & $-0.071 \pm 0.002$ & $-0.096 \pm 0.004$ & & $9.1 \pm 0.2$ & $11.467 \pm 0.007$ & $0.004 \pm 0.008$ & $-0.118 \pm 0.004$\\
        $1.0$ & $6.4 \pm 0.2$ & $11.97 \pm 0.02$ & $-0.063 \pm 0.002$ & $-0.089 \pm 0.002$ & & $7.8 \pm 0.2$ & $11.54 \pm 0.01$ & $0.01 \pm 0.01$ & $-0.120 \pm 0.005$ \\
        $1.5$ & $5.8 \pm 0.4$ & $11.6 \pm 0.2$ & $-0.061 \pm 0.004$ & $-0.075 \pm 0.003$ & & $6.8 \pm 0.2$ & $11.59 \pm 0.01$ & $0.01 \pm 0.01$ & $-0.125 \pm 0.006$ \\
        $2.0$* & $4.58 \pm 0.01$ &  $10^{12} \, h^{-1}$ & $-0.064 \pm 0.001$ & ---  & & $6.07 \pm 0.08$ & $11.701 \pm 0.005$ & $0.015 \pm0.006$ & $-0.124 \pm0.006$\\
        $3.0$ & $3.97 \pm 0.08$ & $11.70 \pm 0.04$ & $-0.055 \pm 0.004$ & $-0.030 \pm 0.005$ & & $4.99 \pm 0.06$ & $11.987 \pm 0.006$ & $-0.016 \pm 0.004$ & $-0.074 \pm 0.009$ \\
        $4.0$* & $3.49 \pm 0.04$ & $11.87 \pm 0.02$ & $-0.039 \pm 0.003$ & $0.01 \pm 0.01$ & & $4.63 \pm 0.04$ & $10^{12} \, h^{-1}$ & $0.030 \pm 0.003$ & ---  \\
        $5.0$* & $3.31 \pm 0.03$ & $10^{12} \, h^{-1}$ & $-0.018 \pm0.002$ & --- & & $3.75 \pm 0.06$ & $10.81 \pm 0.01$ & $0.032 \pm 0.004$ & $0.081 \pm 0.005$  \\
        $7.0$* & $3.40 \pm 0.09$ & $10^{12} \, h^{-1}$ & $0.008 \pm 0.005$& --- & & $5.0 \pm 0.3$ & $10^{12} \, h^{-1}$ & $0.09 \pm 0.01$ & --- \\
\hline
    \end{tabular}
    \raggedright
    * {\footnotesize For at least one group of simulations (i.e., DMO or hydrodynamical), the best-fit model at these redshifts is a pure power law, therefore there are only two free parameters. The power law is normalised at a mass scale of $10^{12} \, h^{-1} \, \rm M_{\odot}$.}
\end{table*}

On top of the three empirical functions described above, we consider the physically motivated analytical model by \cite{Ludlow_2016}. The model predicts the redshift evolution of the concentration-mass relationship from the collapsed mass histories of DM haloes. The formalism uses the Extended Press-Schechter (EPS) theory and assumes that the characteristic density of DM haloes is proportional to the critical density of the Universe at a given collapse redshift. The proportionality constant is the only free parameter of the model, and needs to be calibrated with N-body simulations. We thus re-calibrate such constant so that we obtain the best-fit \cite{Ludlow_2016} model to the data of our DMO simulations. As we can see in the upper panel of Figure~\ref{fig:cM_fit}, the recalibrated \cite{Ludlow_2016} model provides an excellent match to the data, within the statistical errors. However, we cannot apply the \cite{Ludlow_2016} to the hydrodynamic runs, since the underlying formalism ignores the effects of baryons.

To summarise, all models considered provide a reasonable description of the concentration-mass relationship in the DMO simulations. To rigorously determine which function best captures the information embedded in the data without overfitting, we apply again the AIC, as we did for the hydrodynamic-to-DMO mass ratio in Section~\ref{sec:mass_ratio}. The results of the AIC test at $z=0$ are shown in Table~\ref{tab:AIC}. The broken power law is the model favoured by the AIC in both the DMO and hydrodynamic simulations. These models are significantly preferred with respect to the simple power law even in the DMO run. The \cite{Ludlow_2016} model is ranked lowest according to the AIC. The worse AIC score is mainly driven by the higher discrepancy with the data at the higher-mass end, compared to the broken power law. However, this does not mean that it is an inaccurate representation of the data. Indeed, we reiterate that the AIC assesses the relative performance of different models to match a given data set, and not the absolute goodness of fit. In our case, the all power-law models are empirical fits to best reproduce the data. On the contrary, the \cite{Ludlow_2016} model descends from first-principles considerations on the mass collapse history of DM haloes. While we do tune its only free parameter to best describe our data, the model itself is not designed to specifically reproduce the concentration-mass relationship in a given N-body simulation. In fact, it is remarkable that a semi-analytical model relying on a single free parameter still provides an accurate description of the numerical results over six orders of magnitude in the halo mass.

We repeat our AIC analysis for all snapshots considered in this work. The results can be seen in Figure~\ref{fig:cM_akaike}. Clearly, the broken power law is the most favoured model at most redshifts, both in the DMO and hydrodynamic simulations. For some snapshots, a pure power law is preferred. The smoothly broken power law is never the best model according to the AIC, meaning that adding one extra parameter to smooth the transition between the two legs of the relationship does not add any meaningful information, and is thus better avoided. We exclude the \cite{Ludlow_2016} model from Figure~\ref{fig:cM_akaike} because it performs consistently worse than the other fitting functions considered, and showing its considerably higher AIC score would compromise the legibility of the plot. 

We report the best-fit parameters for the model selected by the AIC at each redshift in Table~\ref{tab:fit_params}. The parameters of the broken power law are defined as follows:
\begin{gather}
    \label{eq:conc_bpl}
    c_{\rm 200c} (M_{\rm 200c}) =
    \begin{cases}
        A \left( \frac{M_{\rm 200c}}{M_{\rm ref}} \right)^{\alpha} & \mathrm{if} \; M_{\rm 200c}\leq M_{\rm ref} \\
        A \left( \frac{M_{\rm 200c}}{M_{\rm ref}} \right)^{\beta} & \mathrm{if} \; M_{\rm 200c} > M_{\rm ref} \\ 
    \end{cases} \; ,
\end{gather}
so that $A$ represents the concentration at the mass scale $M_{\rm ref}$ corresponding to the break of the power law, while $\alpha$ and $\beta$ are the slopes in the two legs of the relationship. The pure power law is a special case of equation~\eqref{eq:conc_bpl}, where $\alpha = \beta$. In this scenario, $M_{\rm ref}$ does not represent a break in the concentration-mass relationship, but simply a pivot mass scale regulating the normalisation. A convenient choice for such scale $M_{\rm ref} = 10^{12} \, h^{-1} \, \rm M_{\odot}$, since haloes of this mass are probed by all simulations considered in this work.

We verified that the best-fit model at each snapshot considered typically matches the measured concentration-mass relationship within $0.01-0.04 \, \rm dex$ ($\sim$2\%-10\%), with higher accuracies generally corresponding to lower masses. On the other hand, the spread of the concentration-mass relationship around the mean (see Figure~\ref{fig:cM_rel}) ranges between 0.12 and 0.23 dex ($\sim$32\%-70\%), depending on halo mass and redshift. Thus, any error between the measured concentrations and the predictions of our fit is much lower than the scatter in the concentration-mass relationship.

One might also be concerned about the fact that we combine data sets from simulations with different mass resolutions in order to determine the best-fit parameters of our models. Indeed, there appears to be an offset of up to $\sim 10\%$ in the normalisation of the concentration-mass relationship when moving to a simulation with a different mass resolution (see Figure~\ref{fig:cM_fit}). This is in qualitative and quantitative agreement with an analogous recent study on the concentration-mass relationship in the IllustrisTNG simulations \citep{Anbajagane_2022}. In the Appendix~\ref{app:convergence}, we explicitly verify that improving the mass resolution by a factor of $8$ for every simulation considered typically introduces a variation between 2\% and 10\% ($0.01-0.04 \, \rm dex$) in the normalisation of the present-day concentration-mass relationship, with the larger relative differences typically impacting higher halo masses. At higher redshift (e.g., $z=4$), the differences shrink down to $3\%-6\%$ ($0.01-0.03 \, \rm dex$). The impact of the numerical resolution on concentration is therefore comparable to, or smaller than, the accuracy of our fit. We thus conclude that our fit is robust.

\subsubsection{Evolution of the concentration-mass relationship}

\begin{figure*}
    \centering
    \includegraphics[width=0.99\textwidth]{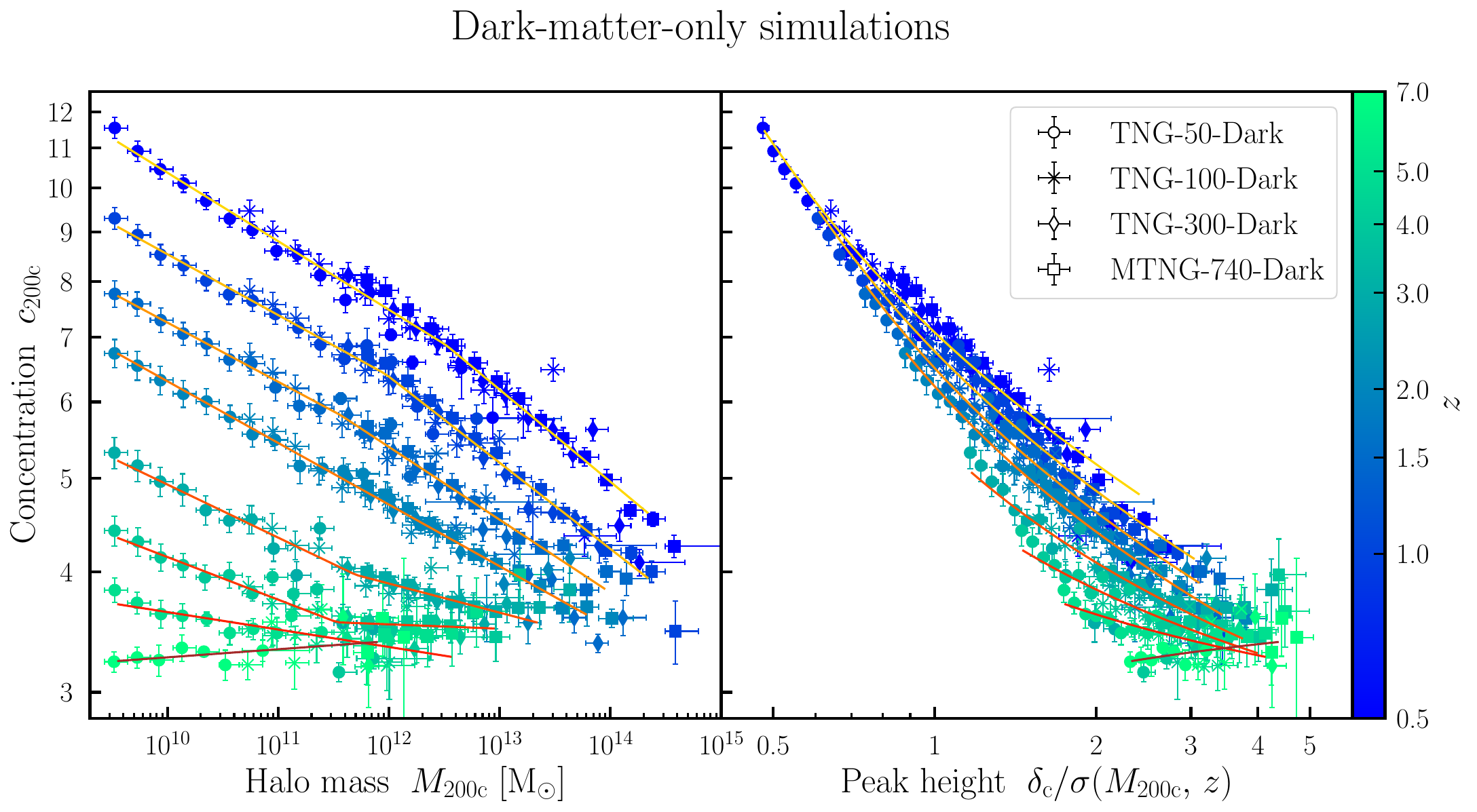}
    \caption{\textit{Left panel}: Concentration-mass relationship for relaxed haloes in the DMO simulations. Data points represent the concentration of the stacked dark matter density profiles of haloes with total mass delimited with the horizontal error bars. The data points are plotted at the median mass within each bin. The vertical error bars show the statistical error deriving from the NFW fit to the stacked profiles. Data points are colour coded according to the redshift considered, as indicated in the colour bar, while their shape refers to the different simulations, as reported in the legend inside the right panel. The solid lines represent the concentration-mass relationship given by best-fit model at each redshift according to the AIC (see Figure~\ref{fig:cM_akaike} and Table~\ref{tab:fit_params}). \textit{Right panel}: Same as the left panel, except that the $x$-axis reports the peak height instead of the total halo mass. The broken power-law or pure power-law models are excellent fits to the data at all redshifts, and across the full halo mass range considered.
    }
    \label{fig:conc_fit_dmo}
    \includegraphics[width=0.99\textwidth]{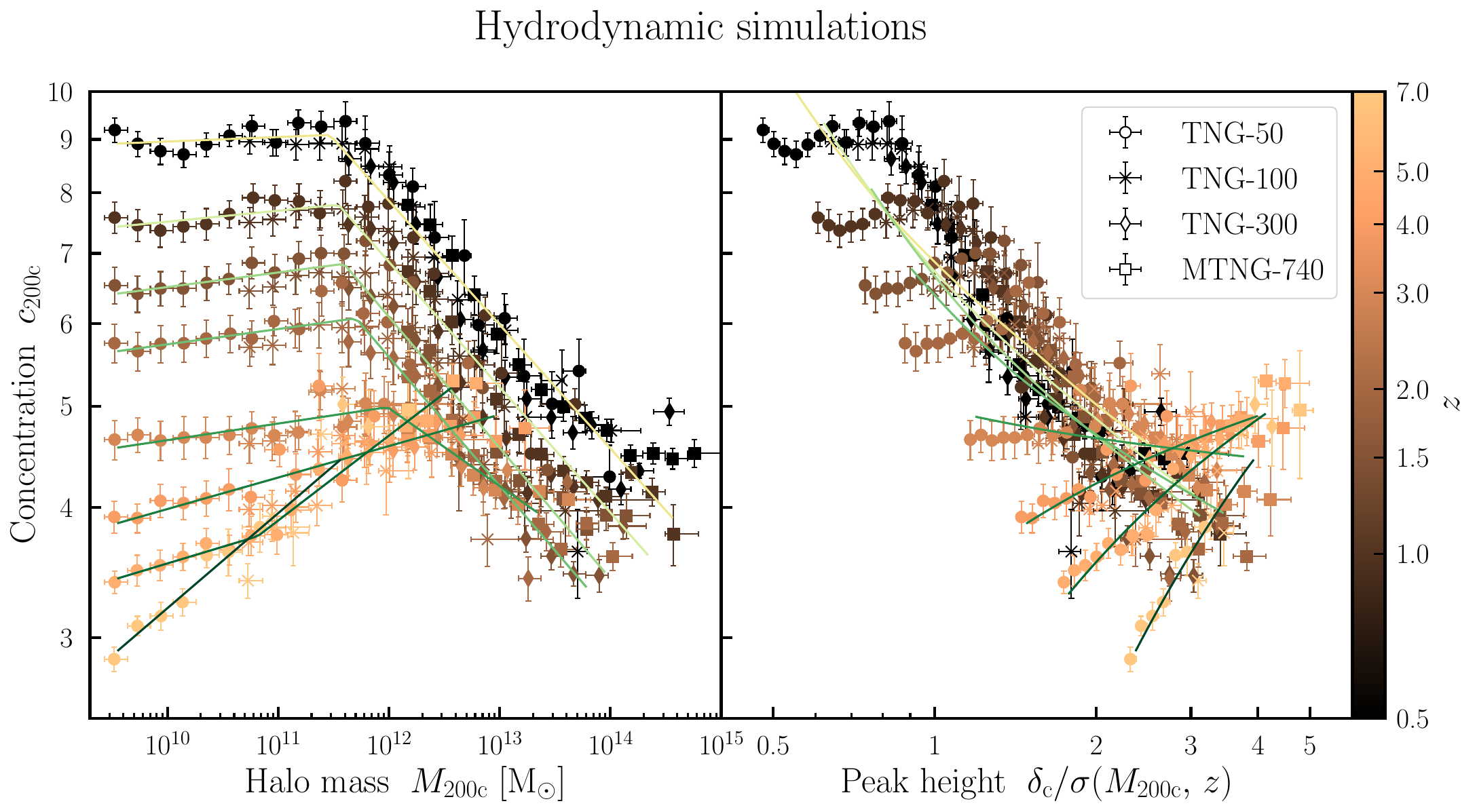}
    \caption{As in Figure~\ref{fig:conc_fit_dmo}, but for the hydrodynamic simulations. The solid lines represent the best-fit functions presented in this work (see Table~\ref{tab:fit_params}) The broken power-law or pure power-law models are excellent fits to the concentration-mass relationship, but fall short of capturing the concentration-peak height relationship at lower masses.
    } 
    \label{fig:conc_fit_fp}
    
\end{figure*}

Having determined the best models representing the concentration-mass relationship in all snapshots, we can now discuss its evolution in the redshift range $0 < z < 7$.

We begin with the DMO simulations, showing their concentration-mass relationships in the left panel of Figure~\ref{fig:conc_fit_dmo}. For each snapshot, we perform exactly the same analysis as described in Section~\ref{sec:conc_z0}. The error bars represent the statistical error on the concentration from fitting the average density profile in each mass bin. As explained in Section~\ref{sec:cM_model}, this generally underestimates the error on the concentration in the higher-mass bin. We verified that the error due to cosmic variance or bootstrapping increases the uncertainty on the concentration to an extent comparable to what we found for $z=0$ (Figure~\ref{fig:cM_fit}). We opt for not including such errors in Figure~\ref{fig:cM_fit} to aid the readability of the plot.

At higher redshift, the normalisation of the concentration-mass relationship decreases. This means that DM haloes of a given mass are less concentrated at higher redshift, since DM had less time to accrete onto haloes and cause further collapse due to self-gravity. The slope of the concentration-mass relationship is less steep at earlier times, and almost flat (if not mildly increasing) at $z=7$. This suggests that DM haloes tend to start off with the same concentration. As time goes by, they collapse under their own gravity. Halo mergers can then generate more massive structures, which will virialise again after a certain relaxation time. Recalling that we are only considering relaxed haloes, it is apparent that higher-mass haloes have had less time to attract DM towards their inner regions since their last major merger. Therefore, higher-mass haloes are less concentrated, and introduce the distinct decline in the concentration-mass relationship.

In the right panel of Figure~\ref{fig:conc_fit_dmo}, we plot the same data as in the left panel, but as a function of the peak height rather than the halo mass. The peak height is defined as $\delta_{\rm c}/\sigma(M_{\rm 200c}, \, z)$, where $\delta_{\rm c}=1.686$ represents the critical density fluctuation for collapse, linearly extrapolated \citep{Peebles_1980, Percival_2005}, and $\sigma(M_{\rm 200c}, \, z)$ denotes the fractional variance of matter density fluctuations in linear theory, averaged over spheres enclosing a mass $M_{\rm 200c}$. The mapping between halo mass and peak height is therefore cosmology dependent, and represents an important quantity in the study of structure formation and evolution. We perform the mapping using the fitting formulae provided by \cite{Ludlow_2016}.

We then show the best-fit (broken) power-law models to the concentration-mass relationships at redshift $z\geq 0.5$, as given by the parameters listed in Table~\ref{tab:fit_params}. Such relationships are plotted with the thin solid lines. In the right panel, the fitting functions are obtained by combining the peak height-mass correspondence provided by \cite{Ludlow_2016} with equation~\eqref{eq:conc_bpl}. At all redshifts, the best-fit models do an excellent job of representing the concentration of DM haloes, both as a function of mass and of peak height.

We repeat the analysis on the hydrodynamic runs, and report the results in Figure~\ref{fig:conc_fit_fp}. As in the DMO run, the normalisation of the relationship decreases at higher redshift. Above a mass scale of $M_{\rm 200c}\sim 10^{11.5} - 10^{12} \, \rm M_{\odot}$, more massive haloes are less concentrated. This is again in line with what we observed for the DMO simulations. But for lower masses, the concentration-mass relationship is essentially flat, at least for $z\lesssim 3$. Below $\sim10^{11.5} \, \rm M_{\odot}$, the gravitational collapse is counteracted by the outward pressure introduced by baryon-driven feedback effects, primarily as stellar winds and supernova explosions \citep{Anbajagane_2022}. These feedback processes, together with AGN-driven winds and jets, are present also at higher halo masses, but are overall less effective. As shown by \cite{Anbajagane_2022}, the energy loss due to gas cooling in the TNG-300 simulation is larger than the energy output due to the kinetic AGN feedback mode in cluster-size haloes ($M\gtrsim 10^{14} \Msun$). The increased relative efficiency of cooling, together with the deeper gravitational potential wells, enables the continued collapse of DM towards the inner regions of the halo. As a result, the relative difference between the concentration in the hydrodynamic and DMO runs is smaller at the higher halo mass end \citep{Anbajagane_2022}.

At $z>3$,  the concentration-mass relationship does not simply become flat, as it was the case for the DMO runs. In the hydrodynamic simulations, the slope of the relationship is reversed at such high redshifts: more massive haloes are now more concentrated. This follows from enhanced adiabatic contraction and subsequent star formation occurring in the cores of massive haloes, which further drives additional DM towards the centre, thereby increasing the concentration. We will support this interpretation in Section~\ref{sec:discussion}, where we will show DM, gas and stellar density profiles within haloes of different mass and at different redshift.

The dependence of the concentration on the peak height exhibits similar differences with respect to the DMO run, as a consequence of the different trend of the concentration-mass relationship. We show the best-fits to the data in both panels of Figure~\ref{fig:conc_fit_fp} obtained from our empirical best-fit models. The formalism successfully captures the main trends observed in the hydrodynamic simulations for the concentration-mass relationship. However, this is not the case for the concentration-peak height relationship. This is not surprising, because the correspondence between halo mass and peak height provided by \cite{Ludlow_2016} was calibrated on DMO simulations, and baryons can break a one-to-one relationship between total halo mass and peak height. This effect should become more important at lower redshift, when more feedback channels are active and contribute to the scatter in the hydrodynamic-to-DMO halo mass ratio. Indeed, we observe a better match to our numerical concentration-peak height relationship at higher redshift.

To summarise, we have found a set of formulae that accurately captures the modification of the concentration-mass relationship measured from DMO simulations in the presence of baryonic physics. We have also shown that the DM distribution within haloes is well represented by an NFW profile both in the DMO and hydrodynamic runs. The combination of these results means that our fitting formulae can be used to predict the DM density profiles of haloes over a wide halo mass and redshift range in the context of a realistic galaxy formation model. This provides a way to correct the results of DMO simulations accurately, making it possible to use them to compare and interpret observational data. As an example, lensing and, in particular, galaxy-galaxy lensing \citep{Tyson_1984, Brainerd_1996, dellAntonio_1996}, is sensitive to the overall matter distribution, where it is important to characterise the response of the DM within and around haloes in the presence of galaxy formation processes like feedback; this work provides a way to account for this effect inside haloes to first order.

\section{Discussion}
\label{sec:discussion}

\subsection{Astrophysical implications}
\label{sec:interpretation}

\begin{figure*}
    \centering
    \includegraphics[width=\textwidth]{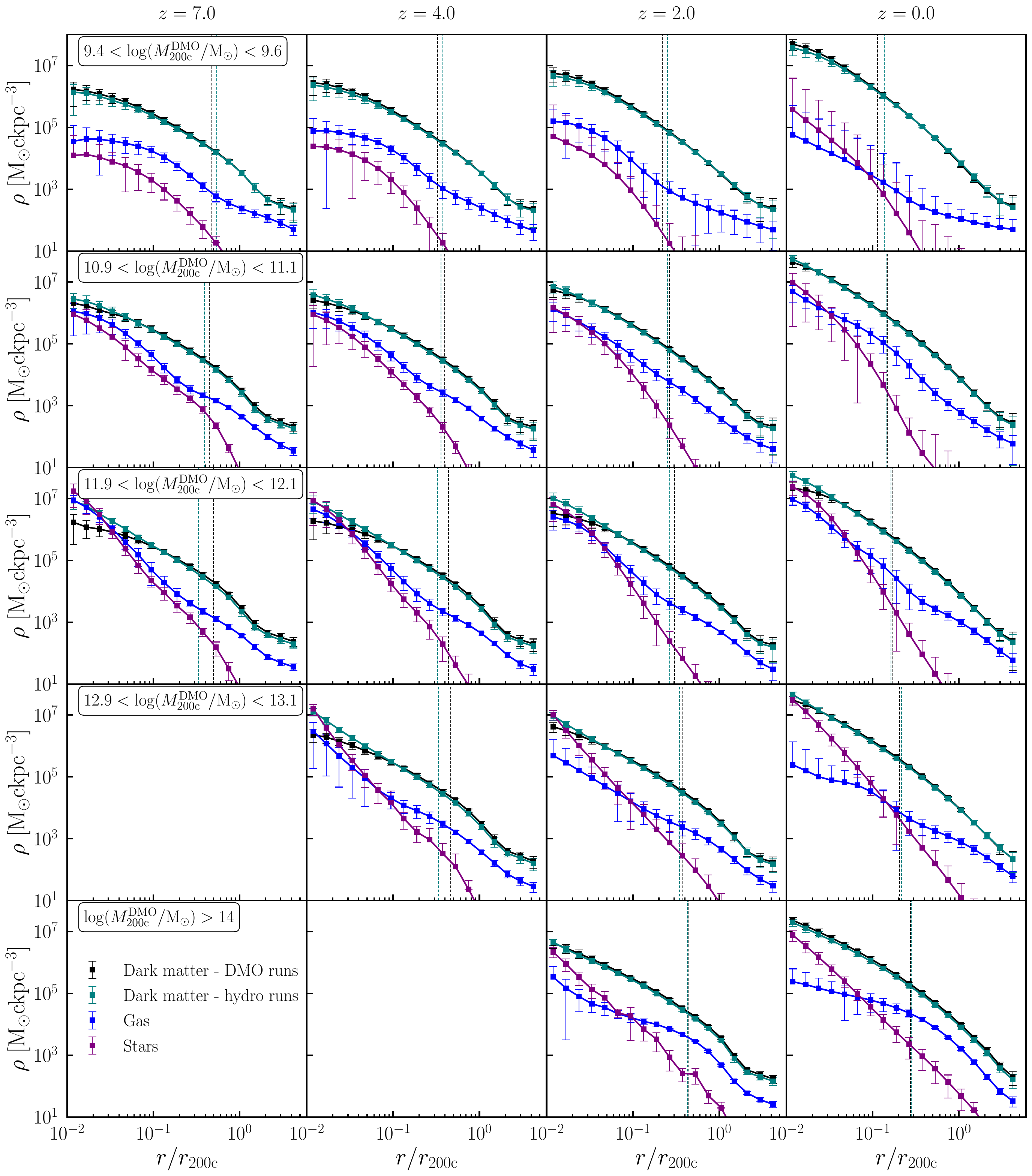}
    \caption{Redshift evolution of density profiles of haloes in all simulations considered. Every row corresponds to a different total halo mass bin in the DMO run, as indicated within the left-most panel. Each column corresponds to a different redshift, as reported above the top panels. The black points represent the comoving density profiles in the DMO run. The teal, blue and purple data sets show, respectively, the comoving density profiles of dark matter, gas and stars in the matched haloes in the hydrodynamic runs. The vertical dashed teal and black lines mark the scale radius for the dark matter density profiles in the hydrodynamic and DMO runs, respectively. At higher redshift and for higher-mass haloes, the gas and stellar density profiles are steeper. Their normalisation does not appreciably change towards redshift, while the dark matter component grows more strikingly by redshift $z=0$. The observed trends explain the redshift-evolution of the concentration-mass relationship in the DMO and hydrodynamic simulations (see Section~\ref{sec:interpretation} for details).}
    \label{fig:all_prof}
\end{figure*}

In Section~\ref{sec:results} we showed how the concentration-mass relationship varies in our simulations when switching from DMO to full hydrodynamic runs. We will now interpret our findings within the context of galaxy formation, focusing on the effects of baryons.

The main conclusion of our analysis is that including baryons in our cosmological simulations flattens the concentration-mass relationship at $M_{\rm 200c} \lesssim 10^{11.5} \, \rm M_{\odot}$. This is not simply caused by numerical artefacts, because we verified that our simulations, which span a wide range of box sizes and mass resolutions, provide consistent results over six orders of magnitude in halo mass (Figures~\ref{fig:cM_rel}-\ref{fig:conc_fit_fp}; see also Appendix~\ref{app:convergence}). The effects that we are seeing are therefore a consequence of baryon-driven physics. To investigate this further,  we now simultaneously explore the distribution of the gaseous, stellar, and DM components within haloes.

For consistency with our previous analysis, we match haloes across the DMO and hydrodynamic runs and extract the density profiles as explained in Section~\ref{sec:results}. We show their redshift evolution, for different halo mass bins, in Figure~\ref{fig:all_prof}. At $z\geq 4$, the density profiles of the gaseous and stellar components are steeper in haloes with mass $M_{\rm 200c} \gtrsim 10^{12} \, \rm M_{\odot}$, especially within 10\% of the virial radius. This is a consequence of the stronger gravitational potential due to the higher mass, which facilitates gas accretion. The accreted gas receives a smaller amount of energy from AGN-driven outflows at $z\geq 4$, because the black hole accretion rate declines steeply with increasing redshift before $z=4$ in the AGN feedback model implemented in the IllustrisTNG simulation \citep{Weinberger_2017}. Thus, at $z\geq 4$, the collapsed gas cools down efficiently via adiabatic contraction, and this favours the production of stars, which are the dominant component within 2\% of the virial radius for higher-mass haloes at $z\geq 4$. The combined abundance of gas and stars in the innermost regions of such haloes further deepens the gravitational potential well, thus attracting dark matter further towards the centre. It then follows that haloes are more concentrated in the hydrodynamic simulations than in their DMO counterparts.

As we can see in Figure~\ref{fig:all_prof}, for $M_{\rm 200c} \gtrsim 10^{12} \, \rm M_{\odot}$ and $z\geq 4$, the DM density profiles in the hydrodynamic simulations indeed appear to be more cuspy than in their DMO counterparts. This is true also at $z=2$, although the effect is less conspicuous than at higher redshift. For $M_{\rm 200c} \approx 10^{12} \, \rm M_{\odot}$, we can see that the scale radius is still smaller for the hydrodynamic run, but there is no significant difference with respect to the DMO variant at higher masses. This reflects the fact that, after peaking at $z\approx 4$, the AGN energy output in the TNG galaxy formation model exhibits only a mild decrease until $z=0$ \citep{Weinberger_2017}. Its sustained effect therefore counteracts gas cooling and star formation, hence preventing the halo from concentrating further.

Moving to lower halo masses, the density profiles of all components are flattened within $\sim 10\%$ of the virial radius even at $z=7$. This is especially evident for $M_{\rm 200c} \approx 10^{9.5} \, \rm M_{\odot}$. In this case, the potential well set by the DM halo is shallower and, consequently, gas does not condense as efficiently as in higher mass haloes. The result of this is that the concentration in the DMO and hydrodynamic variants are similar, with the latter being slightly smaller. 

The different distribution of the gaseous and stellar components within haloes of different mass at $z=7$ then explains why the concentration-mass relationship is monotonically increasing in the hydrodynamic simulations, while the concentration exhibits a weaker dependence on the halo mass in the DMO runs. Instead, the redshift-evolution of the concentration is qualitatively the same regardless of the halo mass. At later times, all haloes tend to deplete their baryons due to stellar or AGN feedback processes. Therefore, they become progressively more DM dominated. This can be clearly seen in Figure~\ref{fig:all_prof}: at $z=0$, the relative difference between the DM profiles and the baryonic components (particularly gas) is larger than at earlier redshift. Thus, the effects of baryons on the internal structure of the DM halo is more `diluted' at later times. The first major consequence is that DM haloes at a fixed mass become more concentrated, as the excess of DM favours further collapse towards the centre of the halo. Secondly, the concentration in the hydrodynamic and DMO runs are generally less discrepant at low redshift: indeed, the respective scale radii are much closer, at least for $M_{\rm 200c} \gtrsim 10^{11} \, \rm M_{\odot}$.

In conclusion, the evolution of the density profiles of DM and baryons within haloes of different mass is consistent with the qualitative behaviour of the concentration-mass-redshift relationship in both the DMO and hydrodynamic simulations considered in this work.

\subsection{Comparison with previous work}

In this section, we compare our main results with the findings of previous related works.

We begin with the halo mass ratio between the hydrodynamic and DMO runs (Figure~\ref{fig:mass_ratio}). Once baryons are introduced in the simulations, the total halo mass varies by only a few percent for $M_{\rm 200c} \gtrsim 10^{14}\, \rm M_{\odot}$, but diminishes at lower masses \citep[see also][]{Castro_2021}. At $z=0$ and for $M_{\rm 200c} \approx 10^{9.5}\, \rm M_{\odot}$, the total mass drops by $\sim 20\%$ with respect to the DMO run. We already mentioned in Section~\ref{sec:mass_ratio} that these results are consistent with previous work with the IllustrisTNG simulation \citep{Springel_2018}. Interestingly, they are quantitatively in broad agreement with analogous works in the literature that adopt other simulations as well. For example, in the GIMIC \citep{Gimic} and EAGLE \citep{EAGLE_Schaye2015} simulations, the mass decreases by $\sim 25\%-30\%$ at $M_{\rm 200c}\lesssim 10^{10}\, \rm M_{\odot}$ when baryons are included, while it remains essentially unchanged above $\sim 10^{13.5}\, \rm M_{\odot}$ \citep{Sawala_2013, Schaller_2015}. However, the trend of the mass ratio is qualitatively different, depending on the simulation considered. In the GIMIC simulation, the hydrodrodynamic-to-DMO mass ratio is monotonically increasing with halo mass \citep{Sawala_2013}, while in EAGLE it resembles a smoothed multiple-step function. By contrast, we find sharp transitions between increasing and decreasing trends around two specific mass scales ($\sim 10^{11.5} \, \rm M_{\odot}$ and $\sim 10^{13}\, \rm M_{\odot}$). 

The diverse trends observed in the literature suggest that not only the presence of baryons, but even the exact modelling of baryon-driven astrophysics in different cosmological simulations is crucial in determining the matter content of haloes at different mass scales.  This was clearly shown, for example, in the Simba \citep{Simba_Dave2019} suite of cosmological simulations, which encompasses five different hydrodynamic runs with varying feedback prescriptions. At $z=0$ and $\sim 10^{12} \, \rm M_{\odot}$, AGN feedback introduces variations of up to $\sim 25\%$ in the total halo mass with respect to a run without any feedback prescription, either stellar or black-hole-driven \citep{Sorini_2022}. This is of the same order of the relative differences that we observe in this work. Thus, whenever trying to model baryonic effects on top of the results of DMO simulations, one should always bear in mind the strong model-dependence of even the most basic quantities, such as the total halo mass.

Similar considerations apply to the concentration-mass relationship. For example, \cite{Duffy_2010} showed that the predictions of the concentration of haloes of a given mass in simulations with different supernova and AGN feedback prescriptions can vary up to $40\%$. The internal structure of DM haloes is then dependent on the complex interplay of cosmological structure formation and astrophysical processes (e.g. \citealt{Chua_2017, Chua_2019, Chua_2022, Arora_2024}; but see also \citealt{Waterval_2022}). It is thus no surprise that different groups found consistently different variations in the concentration-mass relationship when comparing hydrodynamic cosmological simulations to their DMO counterparts \citep[e.g.][]{Schaller_2015, Beltz-Mohrmann_2021}. A comprehensive analysis of the imprint of baryonic physics on the concentration-mass relationship was recently undertaken by \cite{Shao_2023}, using the large suite of CAMELS cosmological simulations. The CAMELS project encapsulates the main features of feedback models of widespread state-of-the-art simulations (EAGLE, Simba and IllustrisTNG) in four parameters that represent the `intensity' of different feedback modes. This facilitates the comparison across boxes that follow different prescriptions for baryonic astrophysics. \cite{Shao_2023} showed that the concentration-mass relationship at $z=0$ deviates from a power law when including baryons. In both Simba and IllustrisTNG type of models, the relationship appears to be decreasing until $M_{\rm 200c} \lesssim 10^{13} \, \rm M_{\odot}$, with an inflection point around $M_{\rm 200c} \lesssim 10^{12} \, \rm M_{\odot}$. The IllustrisTNG models exhibit a plateau in the range $10^{11} < M_{\rm 200c}/\mathrm{M}_{\odot}\lesssim 10^{11.5}$, which is perfectly in line with our findings. The extension to lower halo masses present in our work confirms the significance of the flattening of the concentration-mass relationship in the IllustrisTNG galaxy formation model at the lower mass end.

The flattening presented in this work matches the trends observed for the TNG-50, TNG-100 and TNG-300 simulations by \cite{Anbajagane_2022}. They find that this feature appears at $M_{\rm 200c} \approx 10^{11.5} \Msun$, and extends down to $M_{\rm 200c} \approx 10^{9} \Msun$. In this range, the concentration remains steadily around $c_{\rm 200c} \approx 10$, in line with our results. \cite{Anbajagane_2022} also find that the concentration varies by up to $\sim25\%$ with respect to the DMO versions of the IllustrisTNG runs considered. At intermediate masses, around the point of flattening of the concentration-mass relationship, the concentration increases in the hydrodynamic runs with respect to the DMO variants, but it generally decreases at the lower and higher mass ends. The results of both \cite{Anbajagane_2022} and our work are qualitatively in agreement with the earlier work by \cite{Lovell_2018}, who also found a flattening in the concentration-mass relationship below $M_{\rm 200c} \approx 10^{11.5} \Msun$ (although they used a proxy for the concentration rather than $c_{\rm 200c}$; see their figure 5). The predecessor Illustris simulation also exhibits a break in the concentration-mass relationship, but it occurs at a slightly higher halo mass ($M_{\rm 200c} \approx 10^{12.1} \Msun$). Below this scale, the concentration mildly increases with the halo mass rather than keeping constant \citep{Chua_2017}. The qualitative differences in the relationship between Illustris and IllustrisTNG reflects the adjustments in the underlying feedback models. It is then expected that other hydrodynamic simulations, with completely different feedback schemes, would result in significantly more different concentration-mass relationships \citep[e.g.][]{Schaller_2015}.

\begin{table}
    \centering
    \caption{Power-law fit to the concentration-mass relationship in the DMO simulations considered in this work. The definition of the parameters can be deduced from equation~\eqref{eq:conc_bpl}. We also report the best-fit parameters to the IllustrisTNG-Dark and Illustris-Dark simulations found by \protect\cite{Beltz-Mohrmann_2021}, and to the  EAGLE-DMO simulation \protect\cite{Schaller_2015}, re-normalised to the Hubble parameter $h=0.6774$ and pivot mass scale $M_{\rm ref}=10^{12}\, h^{-1}\, \rm M_{\odot}$ that we have adopted throughout this work.}
    \label{tab:power_law}
    \begin{tabular}{cccc}
        \hline
        $z$ & Model & $A$ & $\alpha$ \\
        \hline
        \multirow{4}{*}{0} & This work & $8.43 \pm 0.03$ & $-0.088 \pm 0.001$ \\
         & IllustrisTNG-Dark & 9.977 & $-0.122\pm 0.005$ \\
         & Illustris-Dark & 8.846 & $-0.125 \pm 0.004$ \\
         
         & EAGLE-DMO & $8.23 \pm 0.16$ & $-0.099 \pm 0.003$ \\ 
         
         & \cite{Dutton_2014} & $8.09 \pm 0.02$ & $-0.101 \pm 0.001$ \\
        \hline
        \multirow{2}{*}{0.5} & This work & $7.14 \pm 0.02$ & $-0.079 \pm 0.001$ \\
         & \cite{Dutton_2014} & $6.56 \pm 0.02$ & $-0.086 \pm 0.001$ \\
        \hline
        \multirow{2}{*}{1.0} & This work & $6.02 \pm 0.02 $ & $-0.075 \pm 0.001$ \\
         & \cite{Dutton_2014} & $5.38 \pm 0.01$ & $-0.073 \pm 0.001$ \\
        \hline
        \multirow{2}{*}{2.0} & This work & $4.59 \pm 0.01$ & $-0.063 \pm 0.001$ \\
         & \cite{Dutton_2014} & $4.121 \pm 0.009$ & $-0.021 \pm 0.002$ \\
        \hline
        \multirow{2}{*}{3.0} & This work & $3.86 \pm 0.02$ & $-0.045 \pm 0.002$ \\
         & \cite{Dutton_2014} & $3.53 \pm 0.03$& $-0.021 \pm 0.002$ \\
        \hline
        \multirow{2}{*}{4.0} & This work & $3.50 \pm 0.02$ & $-0.030 \pm 0.003$ \\
         & \cite{Dutton_2014} & $3.39 \pm 0.03$ & $0.000 \pm 0.003$  \\
        \hline
        \multirow{2}{*}{5.0} & This work & $3.32 \pm 0.02$ & $-0.016 \pm 0.002$ \\
         & \cite{Dutton_2014} & $3.49 \pm 0.05$ & $0.027 \pm 0005$ \\
        \hline
         7.0 & This work & $3.41 \pm 0.07$ & $0.009 \pm 0.005$ \\
         \hline
    \end{tabular}
\end{table}

Comparing different DMO rather than hydrodynamic simulations is more straightforward, as in the absence of baryons, structure formation is driven exclusively by gravity and the expansion of the Universe. The concentration-mass relationship is therefore set solely by the cosmological model. A large body of literature has shown that the concentration-mass relationship in cold DM N-body simulations is monotonically decreasing at $z=0$ \citep[e.g.][]{Duffy_2008, Dutton_2014, Schaller_2015, Beltz-Mohrmann_2021, Uchuu}; this is consistent with our findings here. However, there are quantitative differences regarding the slope and normalisation of the best-fit power law to the present-day concentration-mass relationship. Although we find preference for a broken power law, we also perform a pure power-law fit to our numerical results in order to facilitate the comparison with previous work. We list the best-fit values of the normalisation and slope in Table~\ref{tab:power_law}, following the same definition of the parameters as in equation~\eqref{eq:conc_bpl}. In the same Table, we also report the values obtained in other works. Where a different choice for the pivot mass scale $M_{\rm ref}$ was made, we have corrected the normalisation $A$ to match our own value of $10^{12} \, h^{-1} \, \rm M_{\odot}$.

\cite{Beltz-Mohrmann_2021} found similar slopes for the concentration-mass relationship in the TNG-100-Dark \& TNG-300-Dark simulations and their predecessor Illustris-Dark. However, the normalisation of the relationship in the IllustrisTNG runs is $\sim12\%$ larger, presumably following from the slightly different cosmological model. Compared to our results, \cite{Beltz-Mohrmann_2021} found a higher normalisation and a steeper slope for the concentration-mass relationship in the IllustrisTNG-Dark simulations. This may seem somewhat surprising, given that we adopted the same simulations. However, there are a few crucial differences with respect to our analysis. First of all, \cite{Beltz-Mohrmann_2021} match haloes between hydrodynamic and DMO runs via abundance matching rather than particle IDs. Secondly, we include also the TNG-50-Dark run in our work, which allowed us to extend the analysis to lower halo masses with respect to \cite{Beltz-Mohrmann_2021}. This may impact the parameters of the overall concentration-mass relationship. Finally, we consider only relaxed haloes, whereas \cite{Beltz-Mohrmann_2021} included \textit{all} haloes above $10^{10} \, h^{-1} \, \rm M_{\odot}$. We verified that if we do not restrict ourselves to relaxed haloes, our concentration-mass relationship resembles more closely the findings in \cite{Beltz-Mohrmann_2021}. This comparison confirms that different techniques for extracting the concentration-mass relationship can yield statistically significant differences in the parameters of empirical best-fit functions. It is therefore important to always bear in mind the details of the underlying analysis when comparing the results from different simulations.

Our halo selection criteria and estimation of the mean concentration-mass relationship match those adopted by \cite{Schaller_2015} in an analogous work with the EAGLE simulations. We may therefore expect a closer agreement with their results for the DMO run. However, we must first recall that equation~\eqref{eq:conc_bpl} depends explicitly on the Hubble parameter through the pivot mass scale. Additionally, the Hubble parameter is encapsulated in the definition of the concentration through the virial radius ($c_{\rm 200c} = r_{\rm s} / r_{\rm 200c}$). We thus correct the normalisation parameter found by \cite{Schaller_2015} to match our mass pivot scale and cosmology (the same was done for the Illustris-Dark normalisation reported in Table~\ref{tab:power_law}). Upon such corrections, our normalisation parameter is compatible within one standard deviation with the EAGLE results. We find a less steep slope, which is in slight tension with \cite{Schaller_2015} results. Nevertheless, there is still agreement within three standard deviations. This is reassuring, given the complete independence of the two works.

Both our results and the EAGLE predictions are slightly inconsistent with \cite{Dutton_2014}, who utilised a set of DMO simulations with different box sizes and resolutions \citep{Springel_2005, Maccio_2008, Klypin_2011} to probe the concentration-mass relationship in the mass range $\sim 10^{10}-10^{15} \, \rm M_{\odot}$. They adopted the cosmological parameters from the \cite{Planck14} data release, which are different from the \cite{Planck16} cosmology embedded in the IllustrisTNG and MillenniumTNG simulations. Even if we correct for the different Hubble parameter, as we did for the EAGLE DMO simulation, the discrepancies persist at a statistically significant level. But once again, the details of the analysis undertaken in \cite{Dutton_2014} differ from both \cite{Schaller_2015} and our work. \cite{Dutton_2014} considered haloes with at least 500 particles rather than the more restrictive 5000 threshold imposed in \cite{Schaller_2015} and this work, adopted a slightly different criterion for the selection of relaxed haloes, and a finer binning over a wider range of radial distance when performing the NFW fit. We believe that such differences may introduce systematics that could account for the discrepancies observed.

\cite{Dutton_2014} extend their analysis up to $z=5$, and find that the normalisation of the concentration-mass relationship decreases at higher redshift. Furthermore, the slope of the relationship becomes less steep, and eventually changes sign above $z=4$. Qualitatively, our power-law fits exhibit the same pattern. However, in our case the turning point from an increasing to a decreasing trend of the halo concentration with mass appears at higher redshift, $z>5$. The slope that we measure at $z=7$ is positive, albeit consistent with a flat relationship within less than two standard deviations. These features agree with the findings from the Uchuu N-body simulations \citep{Uchuu}, which also predict a decreasing concentration-mass relationship up to $z=5.2$, and a mildly increasing one at $z=7$. The authors do not provide a power-law fit, but rather utilise a semi-analytical model for the concentration-mass relationship whereby DM halos with low peak height undergo rapid early growth with a universal profile, followed by a slow-growth phase where the halo remains approximately static in physical coordinates \citep{Diemer_2019}. \cite{Uchuu} showed that this model successfully reproduces their numerical results within 5\%.

\cite{Dutton_2014} tested several analytical models for the concentration-mass relationship against their numerical results \citep{NFW, Bullock_2001, Gao_2008, Zhao_2009, Prada_2012}, and concluded that their power-law fits provided a more accurate agreement with the simulated concentration-mass-redshift relationship. In our work, we verified that a broken power law performs better at most redshifts below $z=4$, according to the AIC. We find that a pure power law is acceptable also for the hydrodynamic simulations at $z\geq 4$, but otherwise the broken power law is necessary to accurately represent the flattening of the concentration at the lower-mass end. In general, the qualitatively different behaviour of the concentration-mass relationship across different hydrodynamic simulations \citep[e.g.][]{Schaller_2015, Ragagnin_2019, Beltz-Mohrmann_2021, Ragagnin_2021, Shao_2023, Shao_2024} underscores how the structure of DM haloes is sensitive to the details of the galaxy formation model.

\section{Conclusions and perspectives}
\label{sec:conclusions}

In this study, we investigated the impact of baryons on the concentration-mass relationship of dark matter haloes in the state-of-the-art IllustrisTNG and MillenniumTNG cosmological simulations, which are equipped with almost identical galaxy formation models. Our suite of simulations encompasses a broad range of volumes and mass resolutions, allowing for a detailed examination of haloes across six orders of magnitude in mass ($M_{\rm 200c} \sim 10^{9.5} - 10^{15.5} \, \rm M_{\odot}$), within the redshift interval $0<z<7$. To the best of our knowledge, these are the widest halo mass and redshift intervals probed by cosmological hydrodynamic simulations in a study on the concentration-mass relationship to date. By comparing hydrodynamic runs to analogous dark-matter-only (DMO) variants, we focused on the impact of baryons on the total mass of haloes and on the redshift evolution of the concentration-mass relationship.

The main conclusions of our work are as follows:
\begin{enumerate}
    \item We matched haloes from the DMO runs with their counterparts in the hydrodynamic simulations, and computed the relative variation of their total mass. We find that, on average, the inclusion of baryons in the simulations does not appreciably vary the halo mass above $M_{\rm 200c} \gtrsim 10^{14} \, \rm M_{\odot}$, while the discrepancy can be as large as $\sim 20\%$ for $M_{\rm 200c} \approx 10^{9.5} \, \rm M_{\odot}$ (Figure~\ref{fig:mass_ratio}). We fit the dependence of the halo mass variation as a function of $M_{\rm 200c}$ for all redshifts considered with multiply broken power laws, and provide the best-fit parameters (Table~\ref{tab:mass_ratio_params}).

    \item The concentration of haloes in the DMO simulations at $z=0$ decreases monotonically with mass. The inclusion of baryons flattens the concentration-mass relationship below a mass scale of $M_{\rm 200c} \sim 10^{11.5} \, \rm M_{\odot}$ (Figures~\ref{fig:cM_rel}-\ref{fig:cM_fit}).
    
    \item The steepness of the concentration-mass relationship decreases at higher redshift for the DMO simulations, becoming almost flat at $z=7$. In the hydrodynamic runs, the concentration increases with mass at $z>4$, and decreases thereafter, while always exhibiting a plateau at lower masses (Figures~\ref{fig:conc_fit_dmo}-\ref{fig:conc_fit_fp}).
    
    \item The trends described above are caused by the increased steepness and normalisation of the gas and stellar density profiles in the inner regions of more massive haloes at high redshifts. This effect is largely due to the adiabatic contraction of infalling gas, which promotes star formation. As a result, the higher baryonic density facilitates further dark matter collapse into the central regions of the DM halo, thereby increasing the concentration (Figure~\ref{fig:all_prof}).
    
    \item We tested several empirical and first-principles analytical models for the concentration-mass relationship in the redshift range $0<z<7$ (Figures~\ref{fig:cM_fit}-\ref{fig:conc_fit_fp}). We have shown, with a rigorous information criterion test, that the best-fit model for the results of the DMO and hydrodynamic runs is a broken power law at most redshift considered. A simple power law is generally sufficient to describe the relationship at higher redshift ($z \gtrsim 4$). Instead, the variation of the concentration of DM haloes in the vast mass range considered strongly disfavours the commonly utilised power-law fit at low redshift. We provide the fitting parameters for our best-fit models (Tables~\ref{tab:fit_params}) and for a simple power law in the DMO run, to aid comparison with previous work (Table~\ref{tab:power_law}).
    
\end{enumerate}

The fitting formulae that we provide for the concentration-mass relationship in the DMO and hydrodynamic runs can be used to readily model the density profiles of DM haloes, under the assumption of an IllustrisTNG/MillenniumTNG galaxy formation model in the Planck-18 cosmology. Thus, our results can improve analytical and semi-analytical halo models, as well as the results of cosmological DMO simulations, by incorporating well motivated baryonic effects. Practical applications include a more accurate interpretation of observations that are sensitive to the internal structure of haloes, such as galaxy-galaxy lensing.

Our results qualitatively agree with the literature. We did not include haloes below $M_{\rm 200c} \sim 10^{9.5} \, \rm M_{\odot}$ owing to stringent requirements on the minimum number of resolution elements that guarantees numerical convergence of the density profiles. Adding zoom-in simulations with an analogous galaxy formation model would enable us to expand our study towards lower-mass haloes, hence gaining further insight on the impact of baryon-driven astrophysics on dwarf galaxies. We plan to address this limitation in future work. Another avenue for further development consists in applying our analysis to other cosmological hydrodynamic simulations with different galaxy formation models, which may predict significantly different effects on the concentration-mass relationship. Such questions certainly merit further exploration.

\section*{Acknowledgements}

DS thanks Simon White, Sergio Contreras and Shaun Brown for helpful discussions, and is grateful for the support from the Post-Covid Recovery Fund of Durham University for essential travel connected to the completion of this work. DS and SB acknowledge funding from a UK Research \& Innovation (UKRI) Future Leaders Fellowship [grant number MR/V023381/1]. LH is supported by the Simons Foundation through the collaboration Learning the Universe. VS acknowledges support from the Excellence Cluster ORIGINS which is funded by the Deutsche Forschungsgemeinschaft under Germany’s Excellence Strategy – EXC-2094 – 390783311, and support from the Simons Collaboration on ``Learning the Universe''. CH-A acknowledges support from the Excellence Cluster ORIGINS which is funded by the Deutsche Forschungsgemeinschaft (DFG, German Research Foundation) under Germany's Excellence Strategy -- EXC-2094 -- 390783311. RK acknowledges support of the Natural Sciences and Engineering Research Council of Canada (NSERC) through a Discovery Grant and a Discovery Launch Supplement, funding reference numbers RGPIN-2024-06222 and DGECR-2024-00144. The authors gratefully acknowledge the Gauss Centre for Supercomputing (GCS) for providing computing time on the GCS Supercomputer SuperMUC-NG at the Leibniz Supercomputing Centre (LRZ) in Garching, Germany, under project pn34mo. This work used the DiRAC@Durham facility managed by the Institute for Computational Cosmology on behalf of the STFC DiRAC HPC Facility, with equipment funded by BEIS capital funding via STFC capital grants ST/K00042X/1, ST/P002293/1, ST/R002371/1 and ST/S002502/1, Durham University and STFC operations grant ST/R000832/1. This work made extensive use of the NASA Astrophysics Data System and of the astro-ph preprint archive at arXiv.org.

\section*{Data availability}

The MillenniumTNG simulations will be made fully publicly available at \url{https://www.mtng-project.org} in 2024. The data underlying this article will be shared upon reasonable request to the corresponding authors.




\bibliographystyle{mnras}
\bibliography{MTNG_haloes}




\appendix

\section{Convergence tests}
\label{app:convergence}

\begin{figure*}
    \centering
    \includegraphics[width=0.485\textwidth]{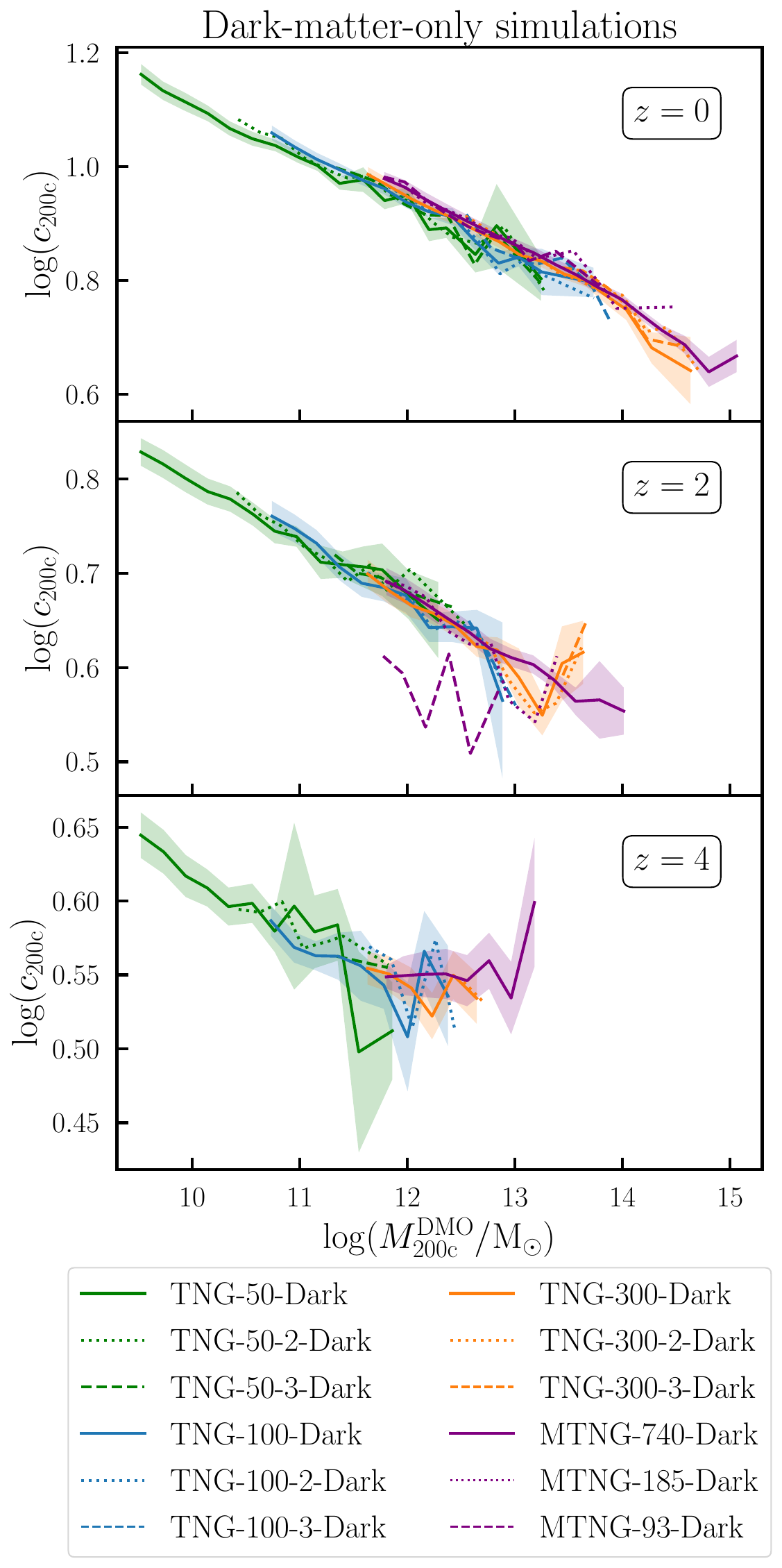}
    \hfill
    \includegraphics[width=0.48\textwidth]{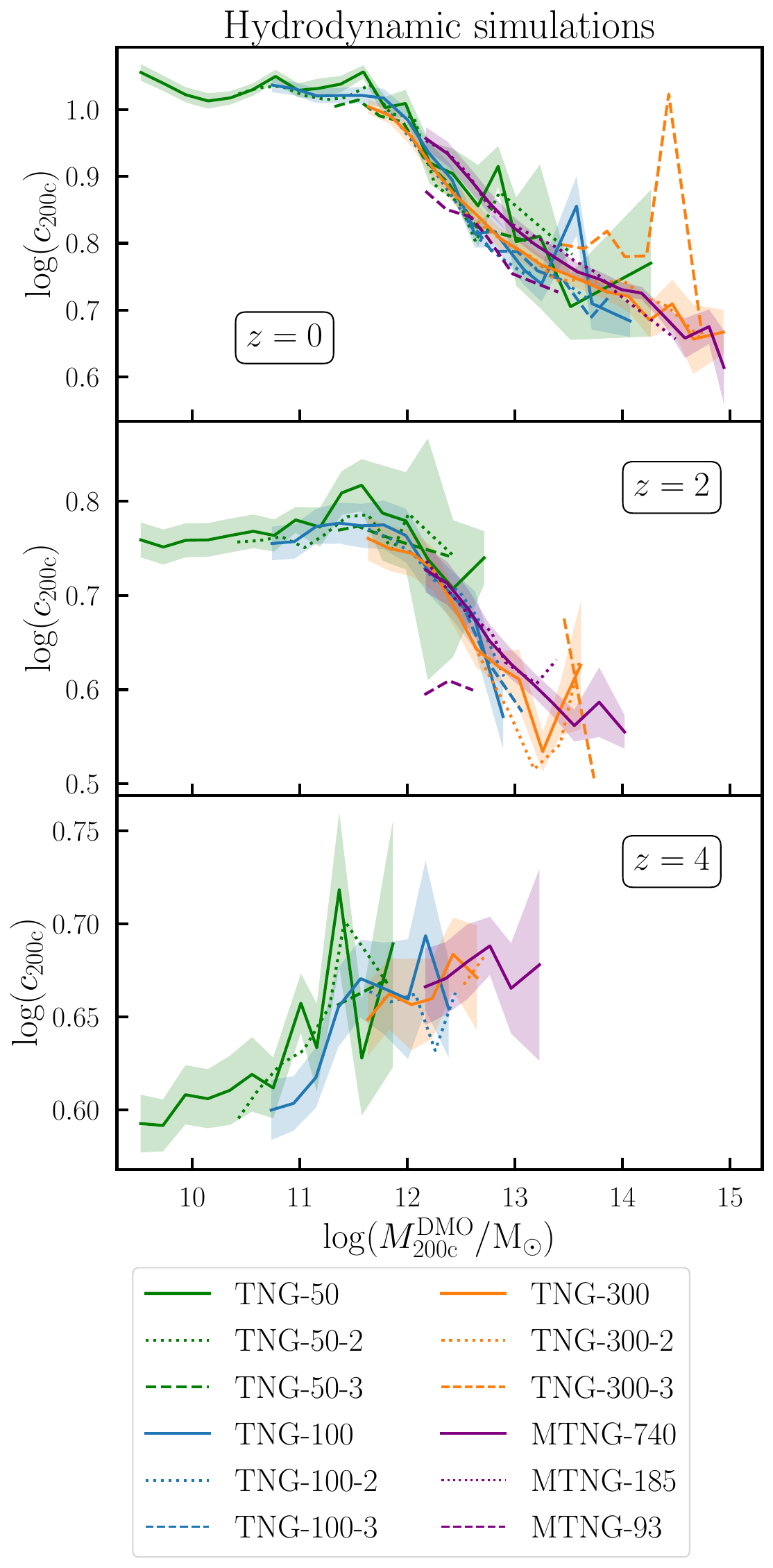}
    \caption{\textit{Left panels}: Convergence test for the concentration-mass relationship predicted by the DMO simulations, at three representative redshifts. For the IllustrisTNG simulations, every colour and line style represent a different box size and mass resolution, respectively, as indicated in the legend at the bottom. For the MillenniumTNG simulation (purple lines), the different line styles correspond to different box sizes. The details of the run corresponding to each simulation label reported in the legend can be found in Table~\ref{tab:simulations}. \textit{Right panels}: As in the left panels, but for the hydrodynamic simulations. For both these runs and their DMO variants, the concentration-mass relationship is well converged with respect to box size and mass resolution, at all redshifts considered.}
    \label{fig:conc_conv}
\end{figure*}

\subsection{Concentration-mass relationship}

In Section~\ref{sec:conc-mass}, we showed that simulations with different box sizes and mass resolutions give consistent results for the concentration-mass relationship across overlapping mass ranges (Figures~\ref{fig:cM_rel}\&~\ref{fig:conc_fit_dmo}-\ref{fig:conc_fit_fp}). In this section, we explicitly test the convergence with respect to the mass resolution for the IllustrisTNG runs. Since we used the MillenniumTNG simulation mainly for extending the upper limit of the halo mass range probed by hydrodynamic simulations, we will test the box-size independence. This is indeed the relevant test for ensuring that our results for clusters and superclusters are not affected by poor statistics.

We show the results of our convergence tests in Figure~\ref{fig:conc_conv}, with left and right panels referring to the DMO and hydrodynamic runs, respectively. We focus on the concentration-mass relationship at present time ($z=0$), cosmic noon ($z=2$), and a suitably high redshift ($z=4$). Every set of simulations is represented with a different colour, as represented in the legend beneath each column of panels. The solid lines are reserved for the fiducial run of each simulation, i.e., TNG-50, TNG-100, TNG-300, MTNG-740, and their respective DMO variants. Other line styles refer to either lower-mass-resolution versions of the IllustrisTNG boxes, or smaller volumes of the MillenniumTNG series. The details of every simulation appearing in Figure~\ref{fig:conc_conv} are reported in Table~\ref{tab:simulations}. The shaded regions represent the maximum among the statistical error on the concentration arising from the fit, cosmic variance, and the bootstrap error, as explained in Section~\ref{sec:conc_z0}. To make the figure more legible, we plot such regions only for the fiducial simulations, although we verified that there is a comparable scatter for the other runs. 

For the IllustrisTNG simulations, the runs with intermediate resolutions match the results of the fiducial runs within the statistical error. Thus, the predictions on the concentration-mass relationship are robust. The convergence is higher for the DMO simulations, while in the hydrodynamic simulations the intermediate-resolution runs can exhibit relatively larger discrepancies. However, the scatter in the hydrodynamic runs is also larger, and generally compatible with the convergence level. Thus, the results obtained from the hydrodynamic simulations are also robust. 

Regarding the MillenniumTNG simulation, convergence with respect to the box size is achieved at $z=0$ for both the DMO and hydrodynamic runs, except for the highest-mass haloes. This is a reflection of the lower statistics in the higher-mass bins following from the cutoff in the halo mass function, and underscores the importance of considering large boxes in order to accurately probe the concentration of superclusters. At $z=2$, the intermediate-volume run exhibits adequate convergence, but the smaller $93 \, \rm cMpc$ box grossly underestimates the concentration-mass relationship. The box size is so limited that no halo satisfies our minimal mass cut of 5000 particles at $z=4$, therefore this run does not appear in the bottom panels.

We note that the concentrations in the DMO runs tends to be slightly biased towards higher values when downgrading the mass resolution of a simulation with a given box size by a factor of 8 from its highest-resolution run (e.g., from TNG-50-Dark to TNG-50-2-Dark, etc.). In the case of the hydrodynamic runs, a lower resolution tends to decrease the concentration of haloes below the mass scale corresponding to a flattening of the relationship ($M_{\rm 200c}\approx 10^{11.5} \Msun$), and to increase concentrations at higher halo masses. In all cases, the relative change in the normalisation of the concentration-mass relationship ranges between $\sim 2\%$ and $\sim 10\%$ for lower-mass and higher-mass haloes at $z=0$, respectively. This is in line with earlier results from an analogous study with the IllustrisTNG simulation by \cite{Anbajagane_2022}. At higher redshift (e.g., $z=4$), the relative differences range between 3\% and 6\%.

To summarise, we proved that we achieve good convergence in the concentration-mass relationship with respect to both mass resolution and volume. Any difference in the normalisation of the relationship due to mass resolution is sub-dominant with respect to the typical accuracy of our best-fit models (see also the discussion in Section~\ref{sec:cM_model}). Therefore, the main conclusions and fitting formulae presented in this work are robust.

\subsection{Halo mass ratio}

\begin{figure}
    \centering
    \includegraphics[width=0.96\columnwidth]{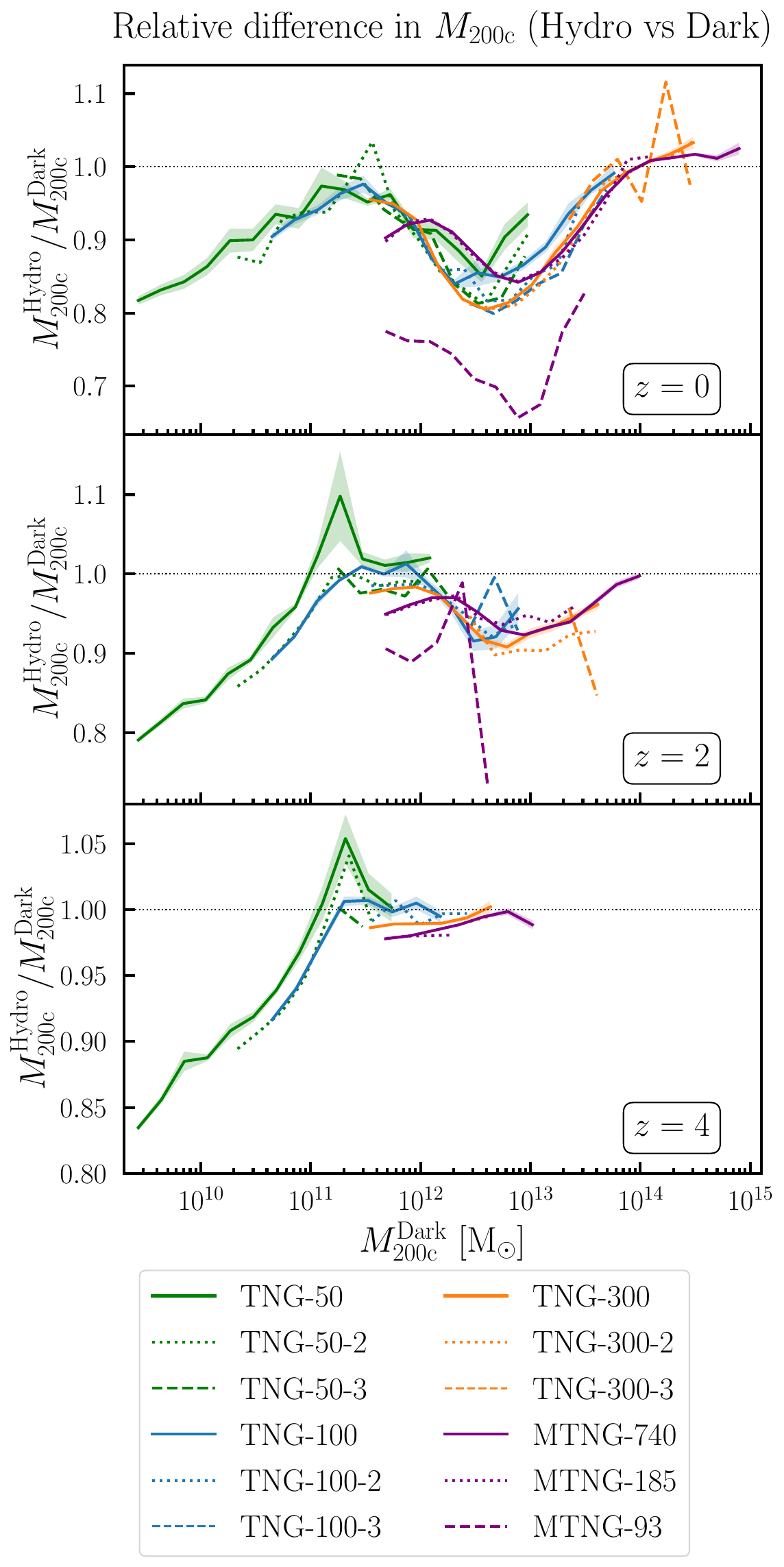}
    \caption{Convergence tests for the hydrodynamic-to-DMO halo mass ratio, as shown in Figure~\ref{fig:mass_ratio}, for three representative redshifts. The details of the run corresponding to each simulation label reported in the legend can be found in Table~\ref{tab:simulations}. The hydrodynamic-to-DMO halo mass ratio is generally converged in the mass range probed by a given set of simulations, but high resolution is crucial to evaluate the ratio at the lowest mass end.}
    \label{fig:mratio_conv}
\end{figure}

We now assess the convergence in the other fundamental quantity that we analyse in this work, i.e., the halo mass ratio between matched haloes across DMO runs and their hydrodynamic counterparts. 

We therefore repeat the same analysis explained in Section~\ref{sec:mass_ratio} on the same simulations considered in Figure~\ref{fig:conc_conv}, and report the results in Figure~\ref{fig:mratio_conv}. The conventions on line styles and colours are the same as in Figure~\ref{fig:conc_conv}. The shaded regions represent the error on the geometric mean for the fiducial runs, but we verified that there is a comparable level of scatter in all other runs. 

The MillenniumTNG simulation exhibits good convergence with respect to the box size at all redshifts. The MTNG-93 box size is again too small to produce reliable results, and heavily underestimates the mass ratio. In this run, only 15 haloes are compatible with our selection criteria at $z=4$, exhibiting a hydrodynamic-to-DMO mass ratio between 0.6 and 0.7. We omit these results from the bottom panel of Figure~\ref{fig:mratio_conv} to make the plot more legible. Clearly, a good statistics of haloes is crucial in order to obtain trustworthy estimates of the hydrodynamic-to-DMO mass ratio. From Figure~\ref{fig:mratio_conv}, we conclude that this is certainly the case for the MTNG-740 run and its DMO counterpart.

The intermediate-resolution IllustrisTNG runs are generally in agreement with the respective fiducial simulations, within the statistical error. The inversions of trend of the hydrodynamic-to-DMO mass ratio consistently occur around the same mass scales ($\sim 10^{11.3} \, \rm M_{\odot}$, $\sim 10^{13} \, \rm M_{\odot}$ and $\sim 10^{14} \, \rm M_{\odot}$) regardless of the mass resolution. Thus, such mass scales have physical significance, and are not merely resulting from numerical artefacts. However, the overall convergence is not as good as in the case of the concentration-mass relationship. At the lower-mass end, the hydrodynamic-to-DMO mass ratio tends to become more sensitive to the mass resolution, especially at higher redshift. This is not unexpected, since haloes of lower mass are represented with a smaller number of particles, and hence more heavily affected by mass resolution.

It is important to note that the slower convergence in mass resolution does not imply that our results are not trustworthy. Indeed, we provided the best-fit functions to the hydrodynamic-to-DMO mass ratio by combining the data from all fiducial simulations together. This means that we can probe the higher-mass haloes with good statistics, thanks to the larger boxes, and at the same time analyse the smaller haloes with the highest mass resolution provided by the smaller simulations. Thus, we always utilise the best data in each end of the expansive mass range that we consider, at every redshift. This ensures the robustness of our results.



\bsp	
\label{lastpage}
\end{document}